\documentclass[twocolumn]{aastex631}

\usepackage{newtxtext,newtxmath}

\usepackage{array}
\usepackage{graphicx}	
\usepackage{amsmath}	
\usepackage{amssymb}    
\usepackage{booktabs}
\usepackage{mathtools}
\usepackage{lipsum} 
\usepackage{xspace}

\newcommand{\tcm}{21$\,$cm\xspace}  

\submitjournal{ApJ}

\begin{document}

\title{Forecasts and Statistical Insights for Line Intensity Mapping Cross-Correlations: A Case Study with 21cm $\times$ [CII]}

\correspondingauthor{Hannah Fronenberg}
\email{hannah.fronenberg@mail.mcgill.ca}

\author[0009-0002-7172-5546]{Hannah Fronenberg}
\affiliation{Department of Physics, McGill University, 3600 Rue University, Montreal, QC H3A 2T8, Canada}
\affiliation{Trottier Space Institute, 3550 Rue University, Montreal, QC H3A 2A7, Canada}

\author[0000-0001-6876-0928]{Adrian Liu}
\affiliation{Department of Physics, McGill University, 3600 Rue University, Montreal, QC H3A 2T8, Canada}
\affiliation{Trottier Space Institute, 3550 Rue University, Montreal, QC H3A 2A7, Canada}

\begin{abstract}
Intensity mapping---the large-scale mapping of selected spectral lines without resolving individual sources---is quickly emerging as an efficient way to conduct large cosmological surveys. Multiple surveys covering a variety of lines (such as the hydrogen \tcm hyperfine line, CO rotational lines, and [CII] fine structure lines, among others) are either observing or will soon be online, promising a panchromatic view of our Universe over a broad redshift range. With multiple lines potentially covering the same volume, cross-correlations have become an attractive prospect, both for probing the underlying astrophysics and for mitigating observational systematics. For example, cross correlating \tcm and [CII] intensity maps during reionization could reveal the characteristic scale of ionized bubbles around the first galaxies, while simultaneously providing a convenient way to reduce independent foreground contaminants between the two surveys. However, many of the desirable properties of cross-correlations in principle emerge only under ideal conditions, such as infinite ensemble averages. In this paper, we construct an end-to-end pipeline for analyzing intensity mapping cross-correlations, enabling instrumental effects, foreground residuals, and analysis choices to be propagated through Monte Carlo simulations to a set of rigorous error properties, including error covariances, window functions, and full probability distributions for power spectrum estimates. We use this framework to critically examine the applicability of simplifying assumptions such as the independence and Gaussianity of power spectrum errors. As worked examples, we forecast the sensitivity of near-term and futuristic \tcm-[CII] cross-correlation measurements, providing recommendations for survey design.

\end{abstract}


\section{Introduction} \label{sec:intro}

During the period of cosmic dawn and and the epoch of reionization (EoR), our Universe's first stars ignited and the first galaxies formed. Emanating from this first generation of galaxies were UV photons that ionized the surrounding neutral hydrogen (HI) over some several hundred million years. The precise timing, duration, morphology, and sources of reionization remains unknown, representing a crucial missing piece in our understanding of our cosmic history.
 
Line intensity mapping (LIM) has the potential to revolutionize our status quo of considerably incomplete information. LIM traces the brightness of specific spectral lines over large three-dimensional volumes, mapping out radial fluctuations using the redshift of a line and transverse fluctuations via angular sky plane information. A judicious choice of spectral lines (balancing practical feasibility with sensitivity to physical processes of interest) opens up heretofore unexplored epochs and scales to direct observation. Particularly promising for the study of the EoR is the \tcm hyperfine transition line of neutral hydrogen, since it is precisely neutral hydrogen that is being ionized and the relevant redshift range is in principle observable \citep{Furlanetto2006Review,Morales&Wyithe,PritchardLoeb2012,LoebFurlanetto2013,10.1088/2514-3433/ab4a73,LiuShawReview2020}.

A number of \tcm observations have been made in the range $0 \lesssim z \lesssim 20$  \citep{10.1088/2514-3433/ab4a73ch8,LiuShawReview2020}. Global signal experiments have targeted the sky-averaged \tcm monopole as a function of redshift. At the high redshifts of cosmic dawn (i.e., during the formation of first stars and galaxies but prior to the systematic reionization of the intergalactic medium), a global signal detection has been claimed by the Experiment to Detect the Global EoR Signature (EDGES) collaboration \citep{EDGES_detection_2018}, although a search for this claimed signal with the Shaped Antenna measurement of the background RAdio Spectrum 3 (SARAS-3) experiment has not yielded a detection \citep{SARAS_non_detection_2022}. At reionization redshifts, global signal experiments have attempted to detect the disappearance of neutral hydrogen as reionization proceeds, and have successfully placed lower limits on the duration of reionzation \citep{2010Natur.468..796B, 2017ApJ...847...64M, 2017ApJ...845L..12S}. Even without a positive detection, such measurements have the potential to set powerful constraints by combining with observations of the kinetic Sunyaev-Zel'dovich in the Cosmic Microwave Background, which tend to place upper limits on the duration \citep{2022PhRvD.105h3503B}.

Complementing global signal measurements are experiments targeting spatial fluctuations in the \tcm brightness temperature. Current-generation instruments typically aim to make a measurement of the power spectrum of these fluctuations, and stringent upper limits have been placed by the Hydrogen Epoch of Reionization Array (HERA; \citealt{HERA_H1C_measurement,2023ApJ...945..124H}), the Giant Meter Wave Radio Telescope (GMRT; \citealt{GMRT_Paciga_2013}), the Low Frequency Array (LOFAR; \citealt{LOFAR_exp, LOFAR_limit_Patil_2017, LOFAR_limit_Gehlot_2019, Mertens_2020_LOFAR_limits}), the Donald C. Backer Precision Array for Probing the Epoch of Reionization (PAPER; \citealt{PAPER_Parsons_2010,PAPER_Cheng_2018,PAPER_Kolopanis_2019}), the Owens Valley Long Wavelength Array (LWA; \citealt{LWA_Eastwood_2019, LWA_Garsden_2021}), the Murchison Widefield Array (MWA; \citealt{MWA_Tingay_2013, MWA_Ewall_Wice_2016, MWA_Beardsley_2016, MWA_Barry_2019, Trott_2020_MWA_limits, MWA_Li_2019, MWA_Yoshiura_2021,2022A&A...666A.106T, MWA_Rahimi_2021,2023ApJ...957...78W}), and the New Extension in Nan\c{c}ay Upgrading LOFAR (NenuFAR; \citealt{2024A&A...681A..62M}). These upper limits have begun to constrain theoretical models \citep{2020MNRAS.493.4728G,2021MNRAS.501....1G,HERA_H1C_2021,2023ApJ...945..124H}, but despite this progress, there remains no positive detection of any spatially fluctuating signal during the EoR. 

The lack of an EoR detection despite such concerted effort is no coincidence; it is a challenging feat. The highly redshifted \tcm line falls in low-frequency radio bands, where there are severe foreground contaminants from more local sources of emission \citep{2005ApJ...625..575S,Wang+2006,2008MNRAS.389.1319J,10.1088/2514-3433/ab4a73ch6}. Examples of such foregrounds include Galactic synchrotron emission, bright radio point sources, Bremsstahlung emission, ionospheric effects, and radio frequency interference (RFI), which dominate over the cosmological signal by up to five orders of magnitude \citep{Liu_Teg_FG}. The effects of these foregrounds must therefore be mitigated, and a vast majority of proposed methods rely on the relative spectral smoothness of foregrounds compared to the more spectrally jagged cosmological signal (see \citealt{LiuShawReview2020} for an overview of methods). While this is a sound idea in theory, in practice instrument systematics can imprint extra spectral structure on the intrinsically spectrally smooth foregrounds, rendering them more signal-like \citep{2020MNRAS.494.3712L}. Therefore, a successful detection of the \tcm signal will necessitate well-calibrated instruments and careful foreground mitigation. These requirements are difficult to execute both to high precision and without being subject to modeling biases. As such, some have looked to use complementary probes to bolster a \tcm detection, while others have called for far more realistic foreground and instrument simulations in order to better understand and analyse real cosmological data. 

One alternative that has seen considerable success in the low-redshift ($z <6$) \tcm cosmology community is to detect the \tcm signal via cross-correlations. On the journey to detecting the low-redshift 21cm power spectrum, a number of \tcm cross-correlations \citep{Masui_HIcross_2013,GBT_eBoss_Wigglez_2021,Parkes_2df_2017,GMRT_HI_stacking_2021, 2023MNRAS.518.6262C, CHIME_HI_stacking_2022,CHIME_Lya} validated the existence of post-reionization \tcm emission in galaxies while providing an easier path towards mitigating systematics such as foregrounds. After about a decade of detecting the low-$z$ 21 cm signal cross-correlation with various other probes, the MeerKAT telescope has recently claimed the first detection of the \tcm auto-spectrum \citep{MeerKAT2023detection}.

In the context of high-$z$ EoR \tcm cosmology, cross-correlations have similarly been touted as promising avenue for systematics mitigation, while also being scientifically interesting in furthering our understanding of the interplay between ionizing sources and the intergalactic medium (IGM; \citealt{Bernal_Kovetz_2022_review}). Possible partners for such cross-correlations are the large suite of emission lines that are being targeted for LIM over a wide variety of redshifts \citep{2017ApJ...835..273G,2021ApJ...911..132Y,2022arXiv220307258K}. These include CO rotational lines \citep{Lidz_CO_2011,2011ApJ...730L..30C,2015ApJ...814..140K,2016ApJ...830...34K,2020ApJ...901..141K,2022JLTP..209..758K,2022ApJ...933..182C,Ihle_2022_COMAP,Lunde_2024_COMAP,2024arXiv240607511S,2024arXiv240607512C,2024arXiv240607861R}, the Ly$\alpha$ line \citep{2013ApJ...763..132S,2014ApJ...786..111P,2018MNRAS.481.1320C,2020PhRvD.101h3032M,2021ApJ...916...22K,2021MNRAS.501.3883R,2022ApJS..262...38L,2022ApJ...931...97K}, the $158\,\mu\textrm{m}$ fine structure line of ionized carbon [CII] \citep{Gong_2011,TIME,2014ApJ...793..116U,2015MNRAS.450.3829Y,2022A&A...659A..12K,2019MNRAS.488.3014P,Chung_2018,2018MNRAS.478.1911P,2019MNRAS.489L..53Y,CONCERTO,2020JLTP..199.1027A,2020arXiv200914340V,2020SPIE11445E..24C,2022SPIE12190E..09E,2022A&A...667A.156B,2023ApJS..264....7C,2023A&A...676A..62V}, and [OIII] \citep{2022MNRAS.515.5813P}, among other proposals such as [OII], HD, H$\alpha$, and H$\beta$ \citep{2017ApJ...835..273G,SPHEREx,2022PhRvD.105h3009B}. Each of these prospects will not only be a powerful probe of astrophysics (e.g., \citealt{2021ApJ...915...33S,2022MNRAS.516.4123M,2022ApJ...933..141P,2022A&ARv..30....5B,2023ApJ...950...40S,2024MNRAS.531.2958H}) and cosmology (e.g., \citealt{2017MNRAS.464.1948F,2019PhRvD.100l3522B,2019PhRvL.123y1301B,2021JCAP...05..067S,2019ApJ...870L...4M,2022ApJ...926..137M,2022JCAP...02..026M,2022arXiv220307258K}) on their own via single-line auto power spectrum measurements, but with the large number of lines available for cross-correlation, multi-tracer techniques will be powerfully constraining \citep{2013ApJ...768...15P,2015aska.confE...4C,2019ApJ...887..142S,2021JCAP...05..068S,2022MNRAS.514.1169A,Maniyar_nulling,2023MNRAS.526.5883S,2023ApJ...950...39M,2023ApJ...957...87R,2023ApJ...958....4L,2024PhRvL.132x1001F,2024PhRvD.109l3518F}.

A particularly interesting prospect is the cross-correlation (and other creative combinations, e.g., \citealt{2018ApJ...867...26B,2019ApJ...874..133B,2023arXiv230800749M}) of the \tcm line and the [CII] $158\,\mu\textrm{m}$ fine structure line \citep{2023MNRAS.523.3503P}. The \tcm-[CII] cross-spectrum $P_{21\rm{cm} \times \rm{[CII]}}(k)$ encodes crucial information as a function of comoving wavenumber $k$, being negative (anti-correlated) on large scales and positive (correlated) on small scales. A measurement of the cross-correlation would yield constraints on the size of ionized regions, the ionization history, and also help constrain astrophysical parameters such as the minimum mass of ionizing haloes, $M_{\rm{turn}}$ and the number of ionizing photons produced by those haloes \citep{Dumitru,Gong_2011}. Most recently, \cite{Moriwaki_2024} have explored the prospects of studying the redshift evolution of these negatively and positively correlated regions, showing that they provide insight on the heating and ionization histories of the IGM.

While the science case for \tcm--[CII] cross-correlations is strong, it is essential to carefully quantify the statistical properties of what we measure. Consider for example the question of whether foregrounds can be suppressed to a reasonable level simply through cross-correlations. If we consider two lines  $A \equiv s_a + n_a + f_a$ and $B\equiv s_b + n_b + f_b$ each comprised of their respective cosmological signal, $s$, foregrounds, $f$, and noise, $n$, then their cross-spectrum can be expressed as
\begin{eqnarray}
\label{eq:x-spec full}
    \langle\tilde{A}\tilde{B}\rangle &=& \langle(\tilde{s}_a + \tilde{n}_a + \tilde{f}_a) (\tilde{s}_b + \tilde{n}_b + \tilde{f}_b)^*\rangle \nonumber \\
     &=&\langle\tilde{s}_a\tilde{s}_b^*\rangle +\langle\tilde{s}_a\tilde{n}_b^*\rangle + \langle\tilde{s}_a\tilde{f}_b^*\rangle + \langle\tilde{s}_b\tilde{n}_a^*\rangle + \langle\tilde{s}_b\tilde{f}_a^*\rangle \nonumber \\
         && + \langle\tilde{n}_a \tilde{f}_b^*\rangle +\langle\tilde{n}_b \tilde{f}_a^*\rangle + \langle\tilde{n}_a\tilde{n}_b^*\rangle + \langle \tilde{f}_a \tilde{f}_b^*\rangle
\end{eqnarray}
where the tildes denote Fourier transforms. Since lines $A$ and $B$ suffer from different sources of foreground contamination and instrument noise, these contributions will be, on average, uncorrelated with one another and with the cosmological signal. Therefore, Equation~\eqref{eq:x-spec full} becomes
\begin{equation}\label{eq:x-corr simplified}
    \langle\tilde{A}\tilde{B}^*\rangle =\langle\tilde{s}_a\tilde{s}_b^*\rangle. 
\end{equation}
However, Equation~\eqref{eq:x-corr simplified} only follows in the case of infinite ensemble averaging. In reality, one can never measure infinitely many samples of each Fourier mode on the sky and it is therefore expected that any cross-spectrum estimation will suffer from residual systematic effects. Equivalently, while the power spectrum is statistically unbiased in the ensemble averaged limit, the \emph{variance} $\rm{Var}(P_{{\rm 21} \times [CII]})$ of \tcm-[CII] cross spectrum will contain terms proportional to each probe's systematics and foregrounds:
\begin{eqnarray}
\label{eq:var_Pcross}
    \rm{Var}\left(P_{{\rm 21} \times {\rm [CII]}}\right) &=& 2\left(P^S_{{\rm 21} \times {\rm [CII]}} \right)^2 + P^S_{\rm 21}P^S_{\rm [CII]} + P^N_{\rm 21} P^N_{\rm [CII]} \nonumber \\
    && + P^F_{\rm 21}P^F_{\rm [CII]} + P^S_{\rm 21} P^N_{\rm [CII]} + P^N_{\rm 21}P^S_{\rm [CII]} \nonumber \\
    && + 
    P^S_{\rm 21} P^F_{\rm [CII]} + P^F_{\rm 21} P^S_{\rm [CII]} + 
    P^F_{\rm 21} P^N_{\rm [CII]} \nonumber \\
    &&+ 
    P^N_{\rm 21} P^F_{\rm [CII]},
\end{eqnarray}
where each term is a product of power spectra, with subscripts indicating the relevant lines and superscripts indicating the power spectrum of cosmological signal ($S$), foregrounds ($F$), and noise ($N$). For example, the cross-specturm signal $P^S_{{\rm 21} \times {\rm [CII]}}$ is equal to $ \langle\tilde{s}_a\tilde{s}_b^*\rangle$. 
The residual systematics can be thought of as either an extra bias that is randomly drawn from these variance terms, or as increased uncertainties on one's final constraints that can (and should) be included in an error budget. Importantly, the form seen here involves \emph{products} of power spectra, which means that non-trivial interactions exist between the different ingredients of one's measurements, making it crucial to include all of them in our explorations.

In this paper we present an end-to-end pipeline for studying LIM cross-correlations. This pipeline takes as input the power spectra of the lines of interest as well as their correlation coefficient (as a function of scale) in order to produce mock cosmological fields. For concreteness, we focus on \tcm $\times$ [CII] cross-correlations. We add a host of simulated foreground contaminants for each line, include the effect of instrumental responses (e.g., their points spread functions; PSFs), model their thermal noise, and finally estimate auto- and cross-spectra. Drawing an ensemble of Monte Carlo realizations (over thermal noise, cosmological fields, and foregrounds) then allows a rigorous quantification of error statistics. In other words, although Equation~\eqref{eq:var_Pcross} motivates and guides our thinking, it is \emph{not} used to generate our results, which contain higher-order statistical information beyond the variance of our observables. We use this technology to investigate the effects of various systematic contaminants and the interplay between them. Along the way, we critically examine various commonly used simplifications in forecasting methodologies, such as (but not limited to) the assumption of Gaussian errors, or the independence of foregrounds in different parts of the sky. As such, our paper is highly complementary to (and builds on) \citet{roy2023crosscorrelation}, which conducted a series of detailed forecasts for LIM cross-correlations. \citet{roy2023crosscorrelation} focused on signal-to-noise and bias statistics, while we also consider non-Gaussian error distributions and ancillary error properties such as power spectrum error covariances and window functions (defined in Section~\ref{subsec:window_functions}). Additionally, whereas we consider the \tcm line and therefore must quantify the attendant complications of interferometry, \citet{roy2023crosscorrelation} consider a broader array of millimeter/sub-millimeter lines. Together, our frameworks can aid in the design of future experiments for cross-correlation, optimizing their sensitivity in the face of systematics. Indeed, as a worked example in this paper we perform a forecast for hypothetical current- and next-generation \tcm-[CII] cross-correlations.

The rest of this paper is organized as follows. In Section~\ref{sec:Motivation_PipelineOverview} we provide an overview of our pipeline and in Sections~\ref{sec:cosmo_fields} to \ref{sec:analysis} we describe in detail each component of this pipeline and the models involved. Readers that are interested in the final forecasts rather than detailed methodology may wish to skip to Sections~\ref{sec:basicforecast} and \ref{sec:designer}. There, we show the results of mock measurements of \tcm-[CII] cross-spectra (and their accompanying statistical properties) under realistic observing conditions for current-generation and futuristic scenarios. We comment on the challenges and opportunities for pursuing such measurements. Along the way, we investigate the extent to which commonly employed approximations are appropriate, and we summarize the results of these investigations for practitioners of forecasting in Section~\ref{sec:sidequestlessons}. Our conclusions are presented in Section~\ref{sec:Conc}.

\section{Motivation and Pipeline Overview} \label{sec:Motivation_PipelineOverview}

In this work we explore the potential of using LIM-LIM cross-correlations to probe the epoch of reionization. As a case study, we explore cross-correlations between \tcm and [CII] LIM observations, although our simulation framework is general and can be easily adapted to explore most LIM-LIM pairs.

The ionized carbon fine structure line, [CII], is of notable interest because it is spatially anti-correlated with the \tcm line. Carbon is first produced in our universe by Population III (Pop III) stars, the first generation stars. Carbon has an ionization energy of $11.26\,\textrm{eV}$ which is below that of hydrogen, meaning that neutral carbon can be more easily ionized. Once ionized, either in the interstellar medium (ISM) or in the IGM, [CII] can undergo the spin-orbit coupling transition $\prescript{2}{}P_{3/2} \rightarrow\prescript{2}{}P_{1/2}$, emitting a photon of wavelength 157.7 $\mu$m. This transition can occur by three different mechanisms: collisional emission, spontaneous emission, and stimulated emission \citep{Suginohara_1999,Basu_2004}. Within galaxies, the main mechanism for emission is collision with ionized gas in the ISM. In the more diffuse IGM, radiative processes are primarily responsible for [CII] line emission. The excitations here are due to spontaneous emission and stimulated emission from collisions with CMB photons. While emission of the CII line occurs both in the ISM and the IGM, the spin temperature of CII line in the IGM is indistinguishable from the CMB photon temperature during the EoR. In the ISM, the spin temperature of [CII] is much greater than the CMB photon temperature during the EoR and the [CII] brightness temperature signal can be observed \citep{Gong_2011}. Since the \tcm brightness temperature emanates from the IGM and the [CII] brightness temperature from the ISM, these two signals are spatially anti-correlated on large scales during the EoR. 

Just as with the \tcm line, [CII] also suffers from foreground contaminants. Luckily, its spectrally smooth far infrared (FIR) continuum foregrounds are not as detrimental to an observation as diffuse synchrotron emission is to \tcm. FIR continuum foregrounds are believed to be removable with negligible residuals via spectral decomposition \citep{Yue_2015}. Still, the [CII] line suffers from bright low-$z$ line interlopers. For example, carbon monoxide (CO) molecules residing at low redshifts can undergo spontaneous rotational transitions, emitting photons that redshift into the same observed frequency band as the photon emitted from the [CII] transition at high redshift. Blind and guided masking techniques have been explored to remove voxels of data thought to be dominated by interlopers \citep{Visbal_2011,Gong_2014,2015ApJ...806..209S,Yue_2015,Masking_Breysse,Sun_2018,Masking_silva}. Masking, however, brings about its own host of challenges since it complicates the survey geometry and alters the bias and amplitude of shot noise due to the down-sampling of the intrinsic line-luminosity function \citep{Bernal_Kovetz_2022_review}. Other interloper mitigation strategies include line identification \citep{2015ApJ...806..234K,Moriwaki_2020}, cleaning using external tracers of large-scale structure \citep{JoseDeInterloping}, analysis of redshift space distortions \citep{2016ApJ...825..143L, 2016ApJ...833..242L, yun-ting}, spectral deconfusion, and of course the subject of this paper, cross-correlations \citep{2010JCAP...11..016V,Gong_2011,Gong_2014,roy2023crosscorrelation}.

\begin{figure}[t]
\includegraphics[width=0.5\textwidth]{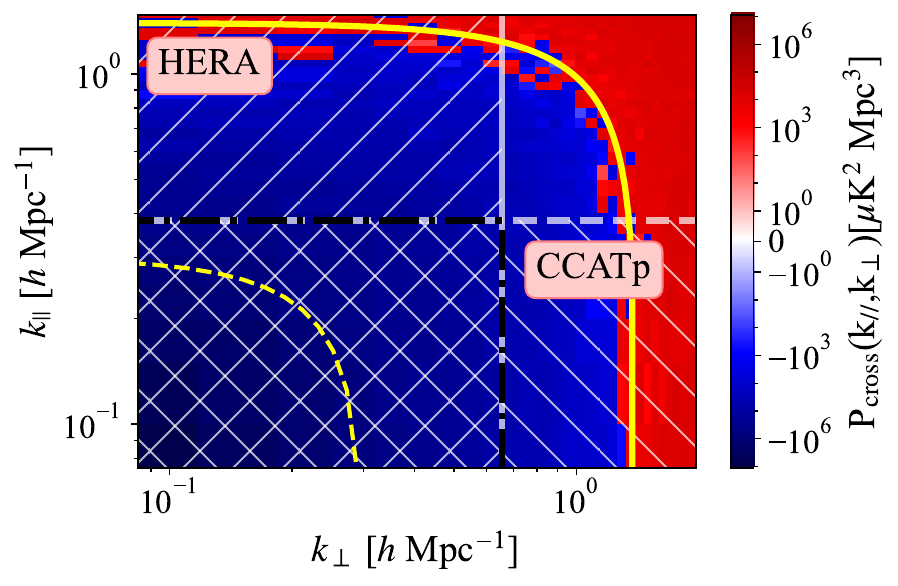}
\caption{Cylindrical 21cm-[CII] cross-spectrum with theory and instrument regions over-plotted. The yellow lines correspond to the theoretically predicted cross-over scale for a fiducial reionization scenario (solid yellow line at $k \sim 1.5$ $h \rm{Mpc}^{-1}$) and for high-mass reionization (dashed yellow line at $k \sim 0.03 $ $h \rm{Mpc}^{-1}$). The modes accessible by HERA and CCAT-prime are boxed off with the jointly accessible modes visible in the cross-hatched region. With only limited Fourier space overlap, cross-correlation measurements may suffer from sensitivity limitations in addition to systematics.}
\label{fig:theory_crosspower}
\end{figure}

While the prospects of measuring a cross-correlation signal are exciting, current-generation experiments are not optimized for joint analyses. In Figure~\ref{fig:theory_crosspower}, the 21cm-[CII] cross spectrum is plotted as a function of $k_\perp$ (spatial wavenumber perpendicular to the line of sight) and $k_\parallel$ (wavenumber parallel to the line of sight) for the fiducial reionization scenario described in \citet{Gong_2011}. As is evident, on large scales the fields are anti-correlated, resulting in negative power shown in blue. On small scales the fields become positively correlated, resulting in positive power shown in red. To help guide the eye, a yellow line has been plotted at the cross-over scale where the power spectrum transitions from negative to positive at $k \sim 1.5\,h \rm{Mpc}^{-1}$. Of course, there remains considerable uncertainty as to how reionization occurred, and as a contrasting case, we also show the cross-over scale of $k \sim 0.03 $ $h \rm{Mpc}^{-1}$ for the high-mass reionization scenario described in \cite{Dumitru}.

In addition to these theory markers, we plot the Fourier modes accessible by current generation \tcm and [CII] instruments, HERA \citep{HERA, HERAII}  and CCAT-prime \citep{2023ApJS..264....7C} respectively. These instruments will serve as our fiducial instruments for the forecasts done in this paper. The modes that are jointly accessible by the two instruments, assuming a $2 \times 2$ square-degree overlap in sky coverage, are shown in the cross-hatched region. These are the only modes for which a cross-spectrum can be estimated with upcoming data and, as is evident, the cross-over scale of the \citet{Gong_2011} fiducial model is far outside the detectable zone. Therefore, it is entirely possible that a near-future measurement may not detect the cross-over scale, although even a null result may rule out certain scenarios (e.g., the \citealt{Dumitru} high-mass scenario).

\begin{figure*}[ht]
\includegraphics[width=1\textwidth]{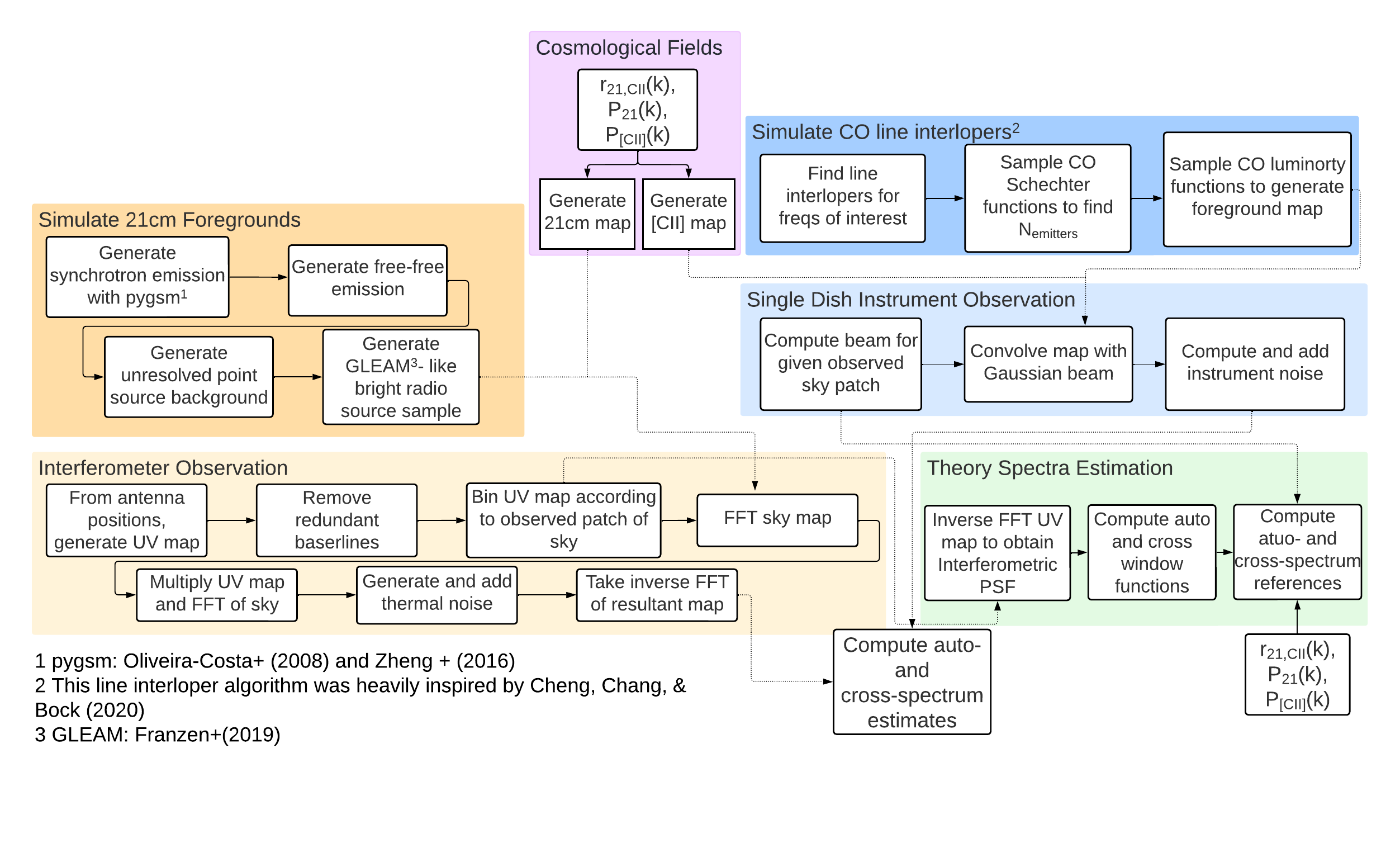}
\caption{Flowchart depicting our end-to-end cross-correlation pipeline. Processes run from top to bottom and left to right. This pipeline produces correlated cosmological fields, simulates 21 cm foreground contaminants, simulates CO line interlopers (i.e. [CII] foregrounds), models both single dish instruments and interferometers, and finally estimates auto- and cross-spectra.}
\label{fig:pipeline}
\end{figure*}

Of course, an overlap in Fourier scales is merely a necessary, not a sufficient condition for a successful cross-correlation measurement. Our simulation pipeline is designed to be an end-to-end effort for a full evaluation of the \tcm-[CII] cross-correlation science yield in the presence of systematics and highly limited Fourier coverage. This pipeline was assembled using our custom software package which we call \textsc{LIMstat}\footnote{\url{https://github.com/McGill-Cosmic-Dawn-Group/LIMstat}}. \textsc{LIMstat} is a statistical framework for the simulation and analysis of line intensity maps. A flowchart of this pipeline can be found in Figure \ref{fig:pipeline}. We start by generating statistically correlated 3D cubes of the \tcm and [CII] brightness temperature fields using our built-in quick correlation model. This produces Gaussian random fields that contain the right fluctuation properties without modeling the physics. These input maps can of course be replaced with boxes from any suite of cosmological simulations. On the \tcm side, Galactic synchrotron emission, bremsstrahlung emission, an unresolved point source background, and bright radio point sources are added to the cosmological signal map. We then simulate an interferometric observation of the final map, given that the vast majority of \tcm cosmology instruments, like HERA, are interferometers. On the [CII] side, CO line interlopers are simulated and it is assumed that continuum emission has been removed. Observation of this interloper-contaminated map by a single dish instrument is then simulated. Finally, these observations are combined to produce cross-spectra.

With the exception of the modelling of Galactic synchrotron emission, which is done using the \textsc{pygsm} package \citep{pygsm_2008,pygsm_2016}, each module of our pipeline was custom built from the ground up for the express purpose of producing realistic simulations of cross-correlations. That being said, the \textsc{LIMstat} software package is modular and the user has the complete freedom to use any component individually depending on the task at hand. In addition, while the surveys used here are HERA and CCAT-prime, our framework is a general one that can be used to model any single dish or interferometric instrument; all of the input parameters can be changed to the design specifications of the instrument one is interested in modelling. In the following three sections, each component of the pipeline is discussed in detail.


\begin{figure*}[ht]
\includegraphics[width=1\textwidth]{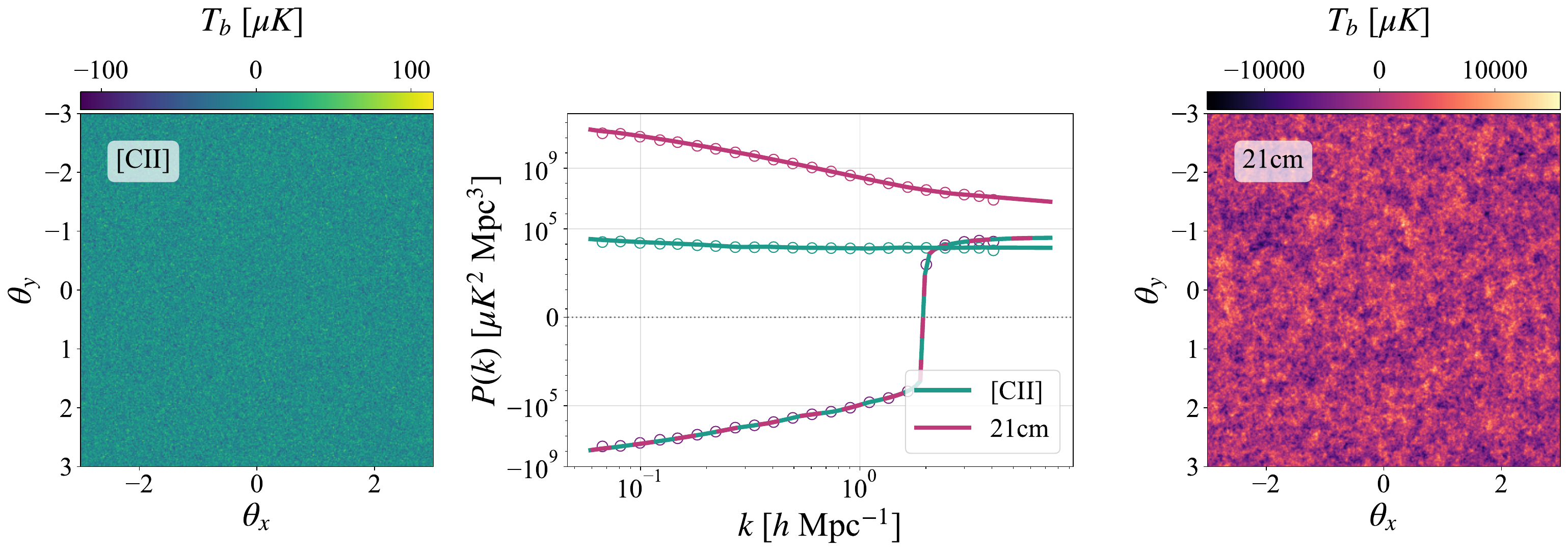}
\caption{Simulated [CII] field (left) and \tcm field (right) over $6 \times 6$ degrees with 1' resolution. In the middle panel, we plot the input theory auto-spectra for \tcm (solid pink) and [CII] (solid turquoise) and also the input theory cross-spectrum (solid alternating pink and turquoise). The spectra estimated from the simulation cubes are plotted with hollow circle markers for \tcm (pink), [CII] (turquoise), and their cross (purple). The correspondence between the spectra from simulated cubes and the input theory spectra validates our approach for generating cosmological realizations with the correct statistical properties.}
\label{fig:Fields+Spectra}
\end{figure*}

\section{Generating Cosmological Fields} \label{sec:cosmo_fields}

In this section, we present how the cosmological signals are modeled in our pipeline. We implement a quick signal simulator that produces maps of two cosmological signals that exhibit accurate cross-correlations. In addition, the individual output maps, on their own, exhibit their respective auto-spectrum statistics. This is purely a statistical model; there is no underlying physics in these simulations beyond what is encoded in the theoretical auto- and cross-spectra of these lines being simulated. This implies, for instance, our simulated fields will not contain any of the non-Gaussianities that are expected in reality. However, we again emphasize that this step of the pipeline can be replaced by more realistic cosmological simulations should parameter estimation be of interest.

 Here we generalize the decorrelation parameter formalism from \citet{Pagano_Liu2020} in order to simulate fields that exhibit the correct scale-dependent correlations. The methodology is as follows. First, we produce the \tcm field as a Gaussian random field with mean zero and power spectrum  $P_{21}(k)$. The Fourier space \tcm field $\tilde{T}_{21} (\mathbf{k})$ is then used to generate the Fourier space [CII] field via the relation\footnote{We follow the standard Fourier convention used in cosmology, as explicitly written out in Appendix~\ref{append:correlation}.}
  \begin{equation}\label{eq:phase_shift}
 \Tilde{T}_{\rm [CII]}(\mathbf{k}) =  f(k)\Tilde{T}_{21}(\mathbf{k}) e^{-i\phi(\mathbf{k})},
\end{equation}
where
\begin{equation}
    f(k) \equiv \sqrt{\frac{P_{\rm [CII]}(k)}{P_{21}(k)}},
\end{equation}
and $\phi(\mathbf{k})$ is a phase that is drawn randomly from a Gaussian distribution with zero mean and standard deviation $\sigma(k)$ (to be specified later). The $f(k)$ factor ensures that the [CII] field will have the [CII] auto-spectrum, $P_{\rm [CII]}(k)$. However, the randomly drawn phase means that when $\Tilde{T}_{\rm [CII]}(k)$ is transformed back to configuration space, its bright and dim regions will have been shifted from their original positions, effecting a (partial) decorrelation of the [CII] field away from the \tcm field.

 Positive and negative correlations are assigned by applying an overall sign flip to Equation~\eqref{eq:phase_shift}, but the degree of correlation or anticorrelation is governed by the parameter $\sigma(k)$. If $\sigma(k) = 0$, every draw of a random $\phi$ in that $k$ bin is 0. The [CII] field then remains perfectly correlated with the 21 cm field in that $k$ bin. On the other hand, if $\sigma(k)$ is large, for each Fourier pixel in that $k$ bin, $\phi$ values are essentially uniformly distributed between $0$ and $2\pi$. This completely randomizes the [CII] field with respect to the 21cm field in that $k$ bin, resulting in uncorrelated fields. In Appendix~\ref{append:correlation} we show that in order for two fields labelled by $a$ and $b$ to have a correlation coefficient
 \begin{equation}
     r_{ab} (k) \equiv \frac{P_{ab}(k)}{\sqrt{P_a(k)P_b(k)}},
 \end{equation}
 one should pick
 \begin{equation}\label{eq:r(k)_sigma}
    \sigma(k) = \sqrt{\ln(|r(k)|^{-2})}.
\end{equation}

In Figure~\ref{fig:Fields+Spectra}, we show an example set of simulated \tcm and [CII] fields with $r(k)$ tuned to match that of \citet{Gong_2011}. This will be the fiducial model that we use throughout the paper. The theory auto-spectra and theory cross-spectrum from \citet{Gong_2011} are plotted with solid curves while the power spectra estimated from the simulation cubes are shown with hollow round markers. As one can see, all of the simulated cubes possess the expected statistical properties.

In summary, our simulation procedure enables the fast generation of paired Gaussian fields that are guaranteed to reproduce chosen auto- and cross-power spectra. The advantage of a fast simulator is that it allows the quick generation of an ensemble of simulations, allowing for a quantification of sample/cosmic variance errors.

\section{Generating Foreground Contaminants} \label{sec:foregrounds}

In this section, we provide an overview of the foreground models used in this pipeline. These are added to the cosmological signals generated in the previous section.

\subsection{$21\,\textrm{cm}$ Foreground Model}
\label{sec:21cmFGs}

\begin{figure*}[t]
\includegraphics[width=1
\textwidth]{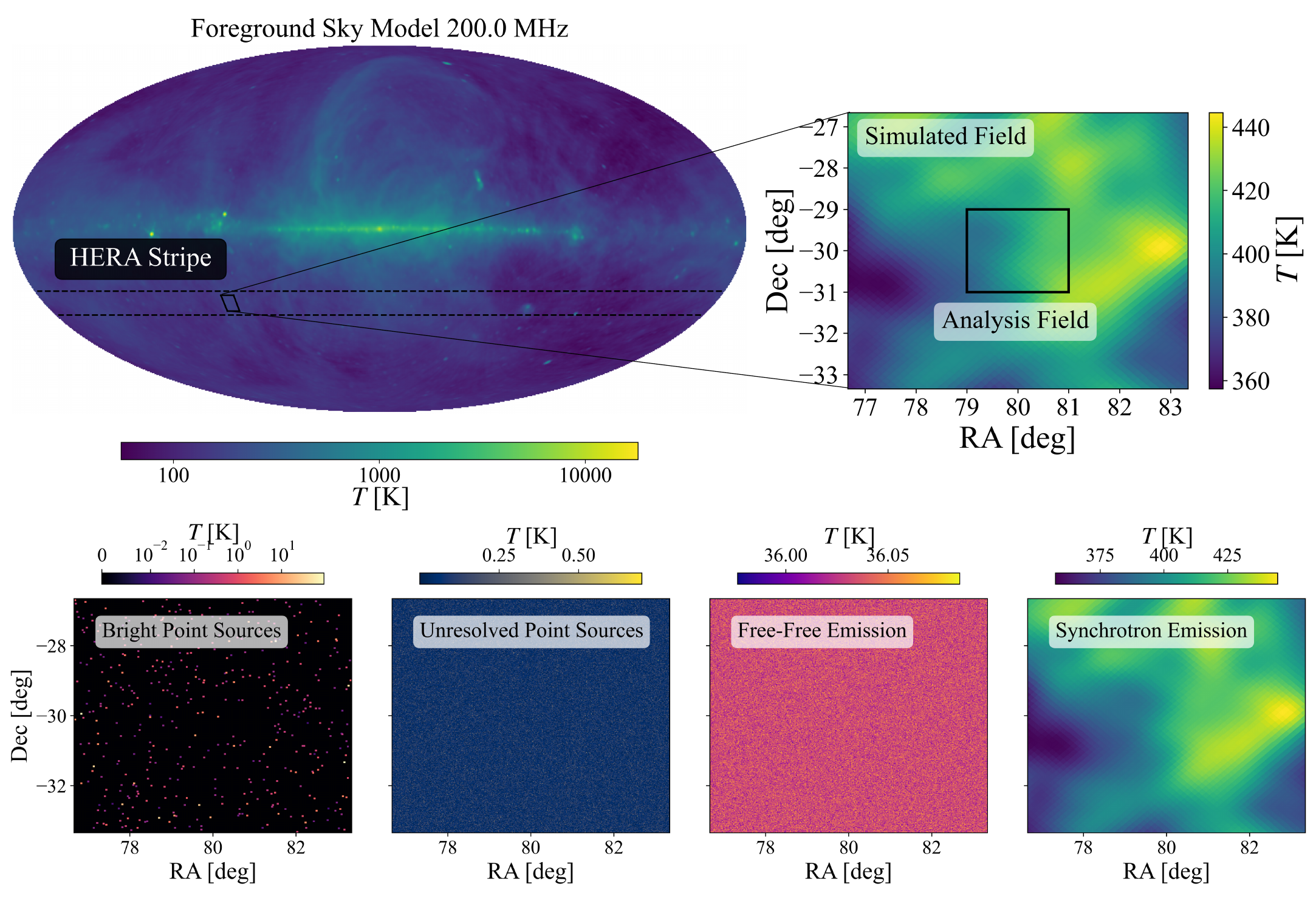}
\caption{Top: Example \tcm foregrounds from our simulation pipeline at $200\,\textrm{MHz}$. The full-sky map shows the Galactic synchtron contribution with the HERA observing stripe superimposed. Each observation is simulated over a $6 \times 6\,\textrm{deg}^2$ simulation patch to capture bright foreground sources caught in low-level sidelobes of an instrument's PSF, in contrast to the smaller $2 \times 2\,\textrm{deg}^2$ analysis patch. Bottom: Example realizations from different constituent components of \tcm foregrounds described in Section~\ref{sec:21cmFGs}.}
\label{fig:21cm_foregrounds}
\end{figure*}

Here we lay out how each of the \tcm foreground contaminants are modeled in the simulation pipeline. In Figure~\ref{fig:21cm_foregrounds}, we show the simulated field as well as each of the \tcm foreground components. Even though HERA and CCATp are expected to overlap over only a small $\sim\!4\,{\rm deg}^2$ patch of sky (the Chandra Deep Field South), we simulate a larger $\sim\!36\,{\rm deg}^2$ patch. The motivation for doing so is that many radio interferometers designed for \tcm have large fields of view (see Section~\ref{subsec:Interferometer}) and point spread functions (PSFs\footnote{Also known as synthesized beams in the radio interferometry parlance.}) that have side lobe structures far away from the central peak. Although these side lobes are generally orders of magnitude lower in amplitude than the central peak, a sufficiently bright point source caught in a side lobe can still leak in from the edges of a field to the center \citep{2016ApJ...819....8P,2022ApJ...938..128X,2023MNRAS.520.4443B}. In order to take this effect into account, we simulate a larger field but perform our cross-correlation analysis on a small analysis field. 

\subsubsection{Galactic Synchrotron Emission}

Galactic synchrotron radiation is the brightest of the diffuse \tcm foreground contaminants. In this work, galactic synchrotron radiation is simulated using the \textsc{pygsm}  package \citep{pygsm_2008,pygsm_2016}. Since there do not exist all-sky maps of our galaxy at all frequencies from observation, this package interpolates over gaps in coverage using a set of principal components that are trained on 29 sky maps between 10 MHz and 5 THz. For the forecasts done here we use the \textsc{pygsm2008}, but \textsc{pygsm2016} is also compatible with this pipeline. 

\begin{figure}[h]
\includegraphics[width=0.5\textwidth]{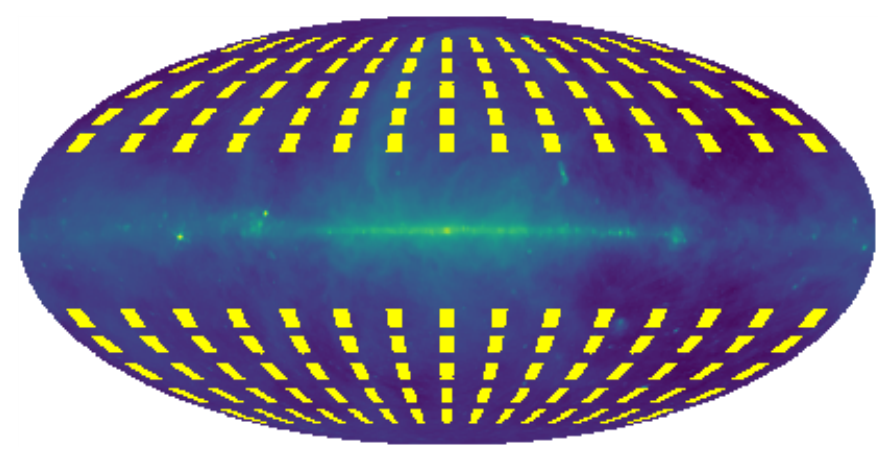}
\caption{Patches of the sky (yellow) from which Galactic synchrotron foregrounds are randomly drawn to form an ensemble of foreground realizations.}
\label{fig:FG_patches}
\end{figure}

Beyond an example map of the synchrotron sky, it is necessary to have a prescription for the statistical distribution from which the map is drawn. This is because one of the goals of this paper is to quantify the increased uncertainty on a power spectrum measurement that is due to foregrounds that survive cross-correlation. Unfortunately, there does not exist a first-principles method to capture the strongly correlated non-Gaussian nature of synchrotron foregrounds. Our approach is therefore to construct an empirical distribution by randomly selecting from different patches of the sky, each of which is the same size as our original simulated field, as illustrated in Figure~\ref{fig:FG_patches}. In fact, this is arguably the more realistic approach, since the strategy of using cross-correlations to mitigate foregrounds is one that implicitly relies on different parts of the sky being different random draws that have different phases in Fourier space. In other words, the relevant probability distribution is the probability distribution of foregrounds in \emph{our} Galaxy, not the probability distribution of a generic synchrotron process in our Universe \citep{2021ApJS..255...26T}.

\subsubsection{Free-Free Emission}

The next contaminant to be modeled is Bremsstrahlung or free-free emission from ionized gas mostly originating from the local ISM. Here we present a simplified model of free-free emission and while this does not capture the morphology, it encapsulates the statistical and spectral properties of this bright diffuse radio foreground. For the free-free emission model, each pixel is assigned a brightness temperature $T$ according to the power law spectrum
\begin{equation}
\label{eq:free-free}
    T(\nu) = A_{\rm{ff}}\left(\frac{\nu}{\nu_*}\right)^{-\alpha_{\rm{ff}}},
\end{equation}
where here, $A_{\rm{ff}} = 33.5$ K and $\nu_* = 150$ MHz. A spectral index $\alpha_{\rm ff}$ is independently drawn for each pixel from a Gaussian distribution, with mean $\overline{\alpha}_{\rm{ff}} = 2.15$ and standard deviation $\Delta \alpha_{\rm{ff}} = 0.01$ \citep{Wang+2006}.

\subsubsection{Radio Point Sources}

Next we consider extragalactic point sources, of which there are two main populations: bright radio point sources and unresolved point sources. It should be noted that a third sub-category also exists, bright extended sources, that is, bright and nearby extragalactic sources that have a spatial extent greater than a single pixel. For \tcm observations in the southern hemisphere, an example of such a source is Fornax A. This sub-category of bright sources is not included in our \tcm foreground models but a treatment of extended sources can be found in \cite{extended_sources}. 

For both populations of point sources, we model a synthetic point source catalog with the same statistics as the Galactic and Extragalactic All-sky MWA (GLEAM; \citealt{GLEAM}) survey based on \citet{Franzen_2016,Franzen_2019}. We draw the number of sources per pixel $n$ from a source count distribution given by
\begin{equation}
\label{eq:SourceCount21cm}
    \log_{10}\left(S^{2.5}\right) \frac{dn}{dS} = \sum_{i=0}^{5} a_i\left(\log_{10}S\right)^i,
\end{equation}
where the parameters $a_i$=\{3.52, 0.307, -0.388, -0.0404, 0.0351, 0.006\} are the best-fit parameters for the GLEAM source count distribution and $S$ is the source flux. This source count distribution is valid for frequencies between $72\,\textrm{MHz}$ and $231\,\textrm{MHz}$ \citep{Franzen_2019}. The source spectra were obtained using a similar expression to Equation~\ref{eq:free-free}, 
\begin{equation}\label{ps_spectrum}
    S(\nu) \propto \left( \frac{\nu}{\nu_*}\right)^{-\alpha_{\rm{ps}}}
\end{equation}
where the spectral index $\alpha_{\rm{ps}} = 0.8$ and the reference frequency $\nu_* = 150$ MHz.

In order to simulate both unresolved and bright extragalactic sources, we simulate two separate point sources maps with different flux limits. For the unresolved point source map, we draw sources with fluxes less than 100 mJy since most peeling techniques can only remove sources whose flux is greater than $\sim\! 10$ to $100\,\textrm{mJy}$. Of course, the exact flux cutoff depends on the resolution and sensitivity of the instrument since, for example, an instrument with lower resolution will smear more sources into the unresolved background. As in \cite{Liu_Teg_FG}, the more conservative bound of $100\,\textrm{mJy}$ is used here. For the bright point source map, we draw sources with fluxes in the range $100\,\textrm{mJy}< S \lesssim   300\,\textrm{Jy}$.

\subsection{[CII] Foreground Model}

Line interlopers are the most problematic foreground contaminant for [CII] observations and this is therefore where we focus our simulation efforts. Carbon monoxide (CO) at low redshifts undergoes spontaneous rotational transitions, emitting a photon that redshifts into the same frequency band as the high-redshift [CII] line. In particular, CO (6-5), CO (5-4), CO (4-3), CO (3-2), CO (2-1), are line interlopers for [CII] observations during the EoR, where the pairs of numbers denote quantum numbers, $J$, indicating the change in the total angular momentum state of the molecule. To simulate all of these CO lines, we follow a similar prescription to \citet{yun-ting}. The main difference with the method employed here is that we approach the modelling from an observational point of view, which has the added benefit of decreasing the computational cost. Instead of building a large and dense lightcone, we instead populate the spectral channels of the instrument with the lines that will have redshifted into that frequency channel. Currently, each channel's interloper map is computed independently; therefore, any line-of-sight interloper correlations are not simulated. In addition, our interlopers do not exhibit any clustering. 

In this foreground model, we populate each pixel with individual CO sources and then draw a luminosity for each source from the Schechter luminosity function of each CO line. The Schechter luminosity function $\Phi$ for CO luminosities $L_{\rm CO}$ is given by 
\begin{equation}\label{eq:OGschechter}
   \Phi(L_{\rm{CO}}) dL_{\rm{CO}}  = \phi_* \left(\frac{L_{\rm{CO}}}{L_*}\right)^{\alpha}e^{-(L_{\rm{CO}}/L_*)} d(L_{\rm{CO}}/L_*),
\end{equation}
and is characterised by the Schechter parameters $\phi_{*}$, $L_{*}$, and $\alpha$. The quantity $\phi_{*}$ is the normalization density, $L_{*}$ is a characteristic luminosity, and $\alpha$ is the power-law slope at low luminosity. The Schechter parameters for the various CO lines as well as the [CII] line used here can be found in \citet{Popping}.

\begin{figure*}[ht]
\includegraphics[width=1\textwidth]{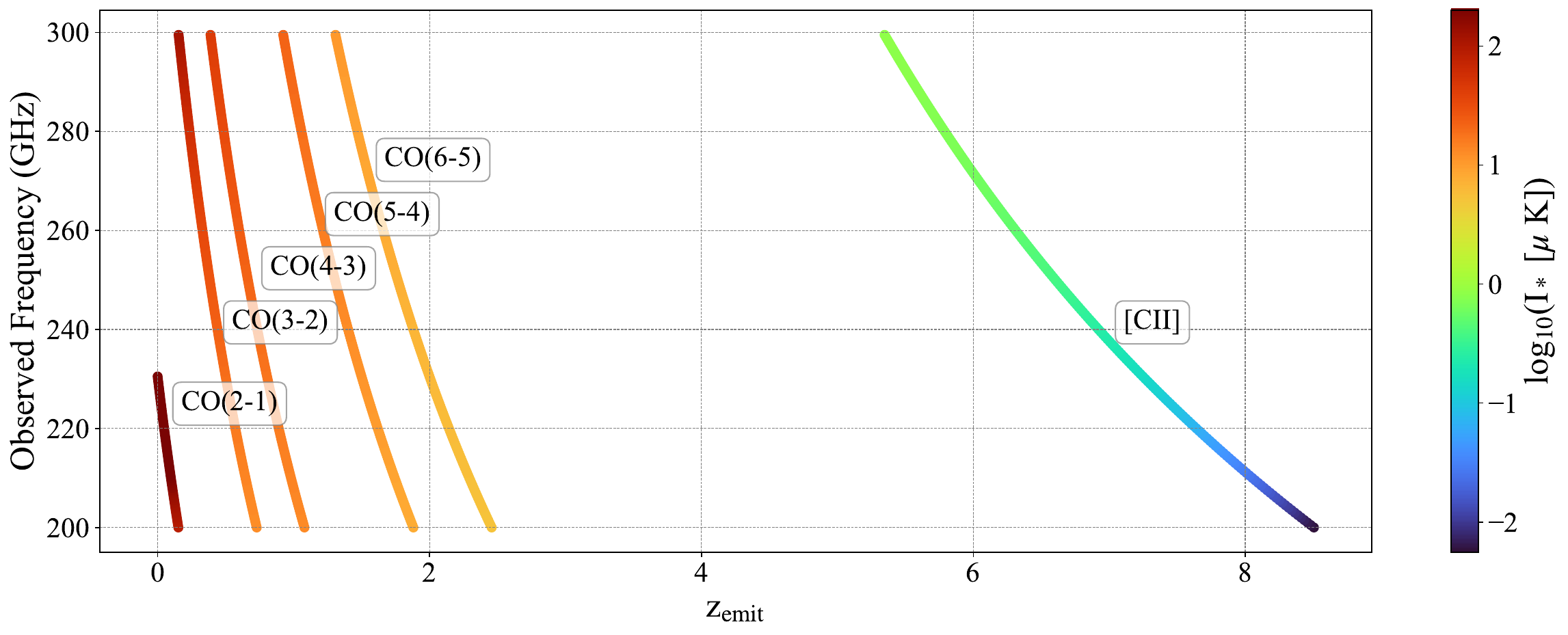}
\caption{Brightness of [CII] and relevant CO emission lines, as functions of emission redshift $z_{\rm emit}$ and observed frequency. At all observed frequencies considered in this paper, CO interlopers will be a strong contaminant to [CII] observations and at the highest redshifts ($z>6$) the CO(2-1) transition is a particularly strong nuisance.}
\label{fig:CO_CII_intensity}
\end{figure*}

We begin by determining which lines redshift into the observed frequency channel and then proceed to interpolate the Schechter parameters for those lines in order to find all of the parameters at $z_{\rm{emit}}$, the redshift at which the line was emitted. Next we proceed to use the Schechter luminosity function to populate the luminosity bins of each line with sources. Having obtained the quantity of interest, $\Phi(L_{\rm{CO}}) dL_{\rm CO}$ given by Equation~\ref{eq:OGschechter}, which is the number of galaxies with luminosity $L_{\rm CO}+dL_{\rm CO}$ per unit volume, we must adjust the number density for the physical size of the voxel that has been simulated. We do this by simply multiplying by the voxel volume 
\begin{equation}
n(L_{\rm{CO}}) dL_{\rm{CO}}  \equiv \Phi(L_{\rm{CO}})dL_{\rm{CO}}V_{\rm{vox}} 
\end{equation}
where $V_{\rm{vox}}$ is
\begin{equation}\label{eq:comovingVol}
    V_{\rm{vox}} = dV_{\rm vox} \Omega_{\rm{pix}},
\end{equation}
with $\Omega_{\rm{pix}}$ being the solid angle subtended by a single voxel in our survey and $dV_{\rm vox}$ being the comoving volume per unit solid angle. The quantity $n(L_{\rm{CO}}) dL_{\rm{CO}}$ is the average number of sources per voxel in each luminosity bin. To ensure there is source count variation from voxel to voxel, we draw the number of sources per luminosity from a Poisson distribution with mean $n(L_{\rm{CO}}) dL_{\rm{CO}}$. The final luminosity is assigned by multiplying the number of sources in each voxel by the luminosity of that bin and then summing over all luminosity bins. Lastly, once the luminosity maps for each line are obtained, they are combined into a single foreground map that has contributions from all line interlopers in a particular channel. In Figure~\ref{fig:CO_CII_intensity}, the characteristic intensity,\footnote{This is obtained through a simple conversion of the characteristic luminosity, $L_*$.} $I_*$, of the various CO lines as well as the [CII] line is shown as a function of redshift and observed frequency. This serves to highlight just how foreground dominated high-$z$ [CII] measurements are.

One potential concern for cross-correlation-based foreground mitigation strategies is the possibility of correlated foregrounds. For example, it is possible that some of the [CII] line interlopers may in fact be emanating from the same galaxies at $0 \lesssim z \lesssim 3$ that are responsible for \tcm point source foregrounds. While this is certainly not impossible, it is not likely. There do exist a handful of hydrogen and deuterium fine structure lines whose rest frequencies are such that if they came from the same $0 \lesssim z \lesssim 3$ galaxies as the line interlopers, they would redshift into the observed \tcm frequency range \citep{ fine_structure_H,deuterium}. That being said, the transition rate of these lines is unknown and the transitions have only been observed under strict laboratory conditions. Still, one may consider broadband emission that could produce these correlation, though we leave such a scenario for future work. For the simulations done here, we generate realizations of radio point source foreground contaminants and CO line interlopers independently.

\section{Simulating Instrument Response}\label{sec:Instruments+Window}

With foregrounds added to our simulated cosmological fields, we proceed to simulating the instruments that observe these foreground-contaminated skies. For concreteness, the conclusions of this paper will be demonstrated using simulations of HERA and CCAT-prime, but we stress that our pipeline is generally applicable to any combination of interferometers and/or single dish experiments.

\subsection{Radio Interferometers}\label{subsec:Interferometer}

In order to simulate both the response and the noise of the HERA array, we compute a set of noisy visibilities. Interferometers measure a primary beam- and fringe-weighted integral of the sky intensity known as the visibility, $V_{ij}$, defined as
\begin{equation}\label{eq:visibilies}
    V_{ij}(\nu) = \int A_{ij}(\mathbf{\hat{s}},\nu) I(\mathbf{\hat{s}},\nu) \exp\left(-2\pi i \frac{\mathbf{b}_{ij}\cdot \mathbf{\hat{s}}
    }{\lambda_{\rm{obs}}}\right)d\Omega
\end{equation}
where $d\Omega$ is the differential solid angle, and $\mathbf{b}_{ij}$ is the baseline vector that characterizes the separation and orientation of the $i$th and $j$th receiving elements (such as dishes). The unit vector $\mathbf{\hat{s}}$ points to the direction of the incoming radiation on the sky. The observing wavelength is denoted by $\lambda_{\rm{obs}}$, $I(\mathbf{\hat{s}}, \nu)$ is the specific intensity, and $A_{ij}(\mathbf{\hat{s}},\nu)$ is the geometric mean between the primary beams of the $i$th and $j$th elements. Equation~\eqref{eq:visibilies} can be compared to a two-dimensional Fourier transform $\tilde{I}$ of the specific intensity, namely
\begin{equation}
    \tilde{I}(\mathbf{u}, \nu) = \int I( \boldsymbol{\theta},\nu) \exp \left( - i 2\pi \mathbf{u} \cdot \boldsymbol{\theta} \right) d^2 \theta,
\end{equation}
where we have invoked the flat-sky approximation to describe positions on the sky in terms of Cartesian angular coordinates $\boldsymbol{\theta} \equiv (\theta_x, \theta_y)$ and have defined a Fourier dual $\mathbf{u} \equiv (u,v)$ to this. One sees that in the limit that the sky is flat and the primary beam is reasonably uniform, each baseline of an interferometer measures a single Fourier mode in the plane of the sky with wavenumbers $u = b_{x} / \lambda_{\rm{obs}}$ and $v = b_{y} / \lambda_{\rm{obs}}$ on the $uv$ plane, where $b_x$ and $b_y$ are the $x$ and $y$ components of $\mathbf{b}_{ij}$, respectively. These modes can then be easily mapped to the comoving wavevector perpendicular to the line of sight, $\mathbf{k}_\perp$, by 
\begin{equation}\label{fourier_baseline}
    \mathbf{k}_\perp = \frac{2\pi \mathbf{b}_{ij}}{\lambda_{\rm{obs}} D_{c}(z_{\rm{emit}})}
\end{equation}
where  $D_{c}(z_{\rm{emit}})$ is the comoving distance to the source emission.
Since our analysis field is much smaller than the area covered by HERA's primary beam (see Table~\ref{tab:HERA}), we omit it in our simulations. In other words, we assume uniform sensitivity over our $4\,\textrm{deg}^{2}$ analysis field. 

\begin{figure}
\includegraphics[width=8cm]{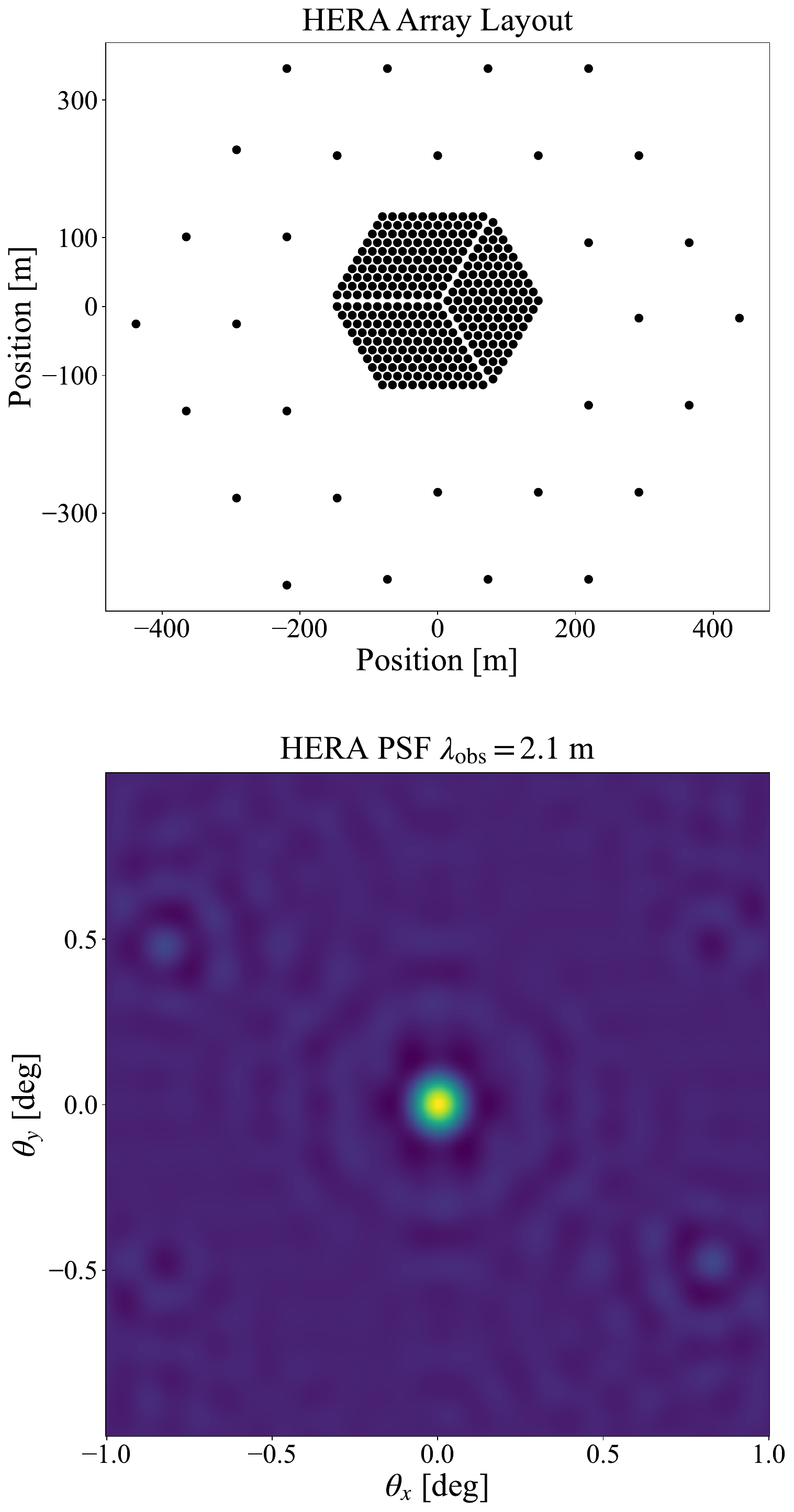}
\caption{Top: HERA array layout assumed for the forecasts in this paper. Bottom: the resulting point spread function.}
\label{fig:HERA_ARRAY}
\end{figure}

To simulate the action of an interferometer, we start by denoting the sky at a particular frequency as $\mathbf{m}$ which stores all of the true pixel intensities on the sky. The measured visibilities then form a map $\mathbf{v}$ in the $uv$ plane given by
\begin{equation}\label{eq:vis_discrete}
    \mathbf{v} = \mathbf{D}\mathcal{F}\mathbf{m}+\mathbf{n}
\end{equation}
where $\mathcal{F}$ denotes the 2D Fourier transform in the angular directions, $\mathbf{D}$ is a binary mask of the $uv$ coverage (recording which $uv$ modes are measured or missed based on the baselines present in an interferometer), and $\mathbf{n}$ is a noise realization in $uv$ space. To generate a noise realization, we populate each pixel in $uv$ space with a complex Gaussian random number with zero mean and standard deviation given by
\begin{eqnarray}\label{eq:HERA_noise}
    \sigma_{\rm{rms}}(u,v) =  \frac{T_{\rm{sys}}}{\sqrt{N_{\rm red} t_{\rm{obs}}\Delta\nu}} \Omega_{\rm{surv}},
\end{eqnarray}
where it is understood that the real and imaginary components are drawn independently and half of the variance is in each component. Here, $T_{\rm{sys}}$ is the system temperature, $t_{\rm{obs}}$ is the total integration time, $\Delta\nu$ is the bandwidth, and $\Omega_{\rm{surv}}$ is the survey area. The quantity $N_{\rm red}$ is the number of baselines that fall into a particular $uv$ cell, and accounts for the fact that HERA is a highly redundant array (see the top panel of Figure \ref{fig:HERA_ARRAY} and \citealt{HERAconfig}) with multiple identical copies of the same baseline. Therefore the noise of any given measurement in the $uv$-plane is reduced by a factor of $1/\sqrt{N_{\rm red}}$. The result of this is that the visibilities of short baselines, which are measured many times over, will generally be less noisy than the visibilities of long baselines that are less redundant.

 The instrument specifications used to generate HERA-like noise are listed in Table \ref{tab:HERA}. It should be noted that in order to match existing HERA noise estimates in the literature, the observing time was reduced by a factor of six. This is to account for the fact that Equation~\eqref{eq:HERA_noise} assumes that all data collected over the duration $t_{\rm{obs}}$ is coherently averaged. In reality, HERA is a non-pointing drift-scan telescope where target fields can only be observed for short durations per sidereal day, and thus the coherent averaging down of noise can only happen in short bursts. Observing different fields will also reduce noise, but only via an incoherent average (at the power spectrum stage). Accounting for this is complicated, given that it crucially depends on simulating rotation synthesis effects and partial redundancy between baselines \citep{2018ApJ...852..110Z}, accounting for subtleties associated with having a drift-scan telescope \citep{LiuShawReview2020}. While this can be done, it would lead to a slight inconsistency with our approach of using different portions of the Galaxy to represent the ensemble properties of possible foregrounds: as Figure~\ref{fig:FG_patches} illustrates, these patches are distributed across the entire sky (excepting the Galactic plane), and in principle would require multiple arrays deployed at different latitudes for all of them to be observable. For simplicity, we thus forgo a detailed rotation synthesis by scaling the integration time to account for the mix of coherent and incoherent integration. We calibrate this scaling so that our code reproduces the \tcm power spectrum sensitivities published in \citet{Pober_2014} when matching their instrumental configurations.

As another approximation, we use a simple fit for $T_\textrm{sys}$ that is given in Table~\ref{tab:HERA}. Since $T_\textrm{sys}$ is partly determined by the receiver temperature ($\sim\!100\,\textrm{K}$) and partly by the sky temperature, it in principle changes when observing the different parts of the sky illustrated in Figure~\ref{fig:FG_patches}. We neglect this variation when computing the noise contribution to our observations, finding that the power-law fit $60\lambda_{\rm obs}^{2.55}$ to be an excellent fit to the mean temperature of our foreground maps. To be clear, this approximation is used only in the determination of noise amplitude in Equation~\eqref{eq:HERA_noise}; the foreground signals that are present in simulated observations are of course specific to each patch of the sky.

 Finally, with our simulated visibilities $\mathbf{v}$, we perform an inverse Fourier transform to obtain our observed HERA map. In radio astronomy parlance, this would constitute a dirty map of the observations, i.e., one where the point spread function (PSF) is not deconvolved from the map. In the bottom panel of Figure~\ref{fig:HERA_ARRAY}, we show HERA's PSF (obtained by Fourier transforming the $uv$ plane's binary mask $\mathbf{D}$). One clearly sees hexagonal features due to the geometry of the array layout. We choose not to remove the effects of the PSF since the blurring effect of the PSF is not a formally invertible operation. Instead, we account for the PSF downstream in our data analysis (Section~\ref{sec:analysis}) via the normalization of our power spectra.

\begin{table}
\setlength{\tabcolsep}{9pt} 
\renewcommand{\arraystretch}{1} 
\centering
\caption{Parameters for the HERA array assumed in this paper. These parameters are based on \cite{DeBoer_2017} and 
\cite{Pober_2014}. **This observing time was further reduced by a factor of six to match existing HERA sensitivity estimates (see Section~\ref{subsec:Interferometer})\label{tab:HERA}}
\begin{tabular}{@{}lllll@{}}
\\ \toprule
Paramters                                         & HERA&&   \\  \midrule
System Temperature, $T_{\rm{sys}}$               & $100+60\lambda_{\rm{obs}}^{2.55} $ K&&   \\
Beam FWHM at 150 MHz, $\theta_{\rm{FWHM
}}$       &8.7 deg&&   \\
Element Diameter      & 14 m &&   
\\
Shortest Baseline      & 14.6 m&&   
\\
Longest Baseline (core)          & 292 m&&   
\\
Longest Baseline (outrigger)          & 876 m&&   
\\
EoR Frequency Range, $\nu_{\rm{obs}}$          & 100-200 MHz&&   \\
Channel Width, $\delta\nu$  & 97.8 kHz&& \\

Integration time on field, $t_{\rm{obs}}$ & 170** hours&&
\\
\bottomrule

\end{tabular}
\end{table}

\subsection{Single Dish Instruments}\label{subsec:single_dish}

The other instrument that we simulate in this paper is CCATp, which is a single dish telescope. This instrument is modeled by convolving the sky with a 2D Gaussian beam with standard deviation equal to the diffraction limited angular resolution, that is, $\theta_{\rm{CCAT}} \sim \lambda_{\rm{obs}}/D$ where $D$ is the diameter of the dish. After the field has been convolved, a noise realization is added. This noise map consists simply of white noise drawn from a Gaussian with mean zero and standard deviation $\sigma_{\rm{rms}}$, which is given by
\begin{equation}
    \sigma_{\rm{rms}} = \frac{\sigma}{\sqrt{t_{\rm{pix}}}}
\end{equation}
where $t_{\rm{pix}}$ is the total integration time per pixel, and 
\begin{equation}
    \sigma = \frac{\sigma_{\rm{pix}}}{\sqrt{\Omega_{\rm{sim}}/
    \Omega_{\rm{pix,CCAT}}}}
\end{equation}
where $\sigma_{\rm{pix}}$ is the quoted per pixel sensitivity of \mbox{CCATp}, $\Omega_{\rm{sim}}$ is the size of a pixel in the simulated cube, and \mbox{$\Omega_{\rm{pix,CCAT}} = \pi\theta^{2}_{\rm{FWHM}}/4\ln 2$} is the nominal pixel size of CCATp observations. Note that despite the confusing notation, $\sigma$ is not a standard deviation. This can be seen by examining its dimensions in Table \ref{tab:ccat}. There, we summarize the instrument specifications for CCATp that are used in this paper, which are based on parameters given in \citet{Breysse_Alexandroff_2019,Chung_2018}.

\begin{table}[t]
\setlength{\tabcolsep}{11pt} 
\renewcommand{\arraystretch}{1} 
\centering
\caption{Parameters for [CII] survey CCATp  assumed in this paper. These parameters are based on \cite{Breysse_Alexandroff_2019}, \cite{Chung_2018}. It should be noted that the parameters with the superscript ($*$) are frequency-dependent quantities and the values in the table were computed at $237\,\textrm{GHz}$.\label{tab:ccat}}
\begin{tabular}{@{}lllll@{}}
\toprule
Parameters                                         &  CCATp   & &   \\  \midrule
System temperature, $\sigma_{\rm{pix}}$ MJy sr$^{-1}$ s$^{1/2}$  & 0.86*        &      &   \\
Beam FWHM, $\theta_{\rm{{FWHM}}}$ (arcmin)        & 0.75*        &      &   \\
Dish Diameter (m)           & 6     & &   
\\
Frequency Range, $\nu_{\rm{obs}}$ (GHz)           & 210-300     &      &   \\
Channel Width, $\delta\nu$ (GHz) & 2.5         &      &   \\
Number of Detectors, $N_{\rm{det}}$                                    & 20          &      &   \\

EoR Survey Area, $\Omega_{\rm{surv}}$ (deg$^2$)             &  4 &     & 
\\ \bottomrule
\end{tabular}
\end{table}

\section{Simulating a Data Analysis Pipeline}
\label{sec:analysis}

After simulating our cosmological signals (Section~\ref{sec:cosmo_fields}), adding foregrounds (Section~\ref{sec:foregrounds}), and putting the resulting sky through simulated instruments (Section~\ref{sec:Instruments+Window}), a crucial last component of an end-to-end analysis is to include a mock data analysis. Nuanced data analysis choices affect one's final uncertainties on a cross-power spectrum and in this section we make our choices explicit.

\subsection{Power Spectrum Estimation}
After Section~\ref{sec:Instruments+Window}, we have our instrument-processed three-dimensional surveys. These are the inputs to our cross power spectrum estimator $\hat{P}_{ab}$. As a first step in our power spectrum estimation, we multiply both surveys by a Blackman-Harris taper function $B$ \citep{harris1978use} in the radial direction, as is often done in \tcm analyses \citep{2013ApJ...776....6T,2016ApJ...825....9T, HERA_H1C_measurement,HERA_H1C_2021,2023ApJ...945..124H}. This is done to prevent harsh non-periodic discontinuities at the edges of one's data cubes, which occur because the \tcm foregrounds are considerably brighter on the low-frequency end of one's survey than the high-frequency end. Leaving such discontinuities in the data causes ringing upon a Fourier transform, which undesirably scatters foreground contaminants from the spectrally smooth modes at low $k_\parallel$ to spectrally unsmooth modes at high $k_\parallel$ that should otherwise be reasonably contaminant-free. Denoting our post-tapering survey data to be $T^{\rm tap} (\mathbf{r})$, we then employ an estimator that is very similar to one that would be used for perfect theoretical data, namely one where the surveys are Fourier transformed, cross-multiplied, and normalized to form 
\begin{equation}
\label{eq:PspecEstimator}
    \hat{P}_{ab} (\mathbf{k}) \equiv \frac{\tilde{T}^{\rm tap}_{a}(\mathbf{k}) \tilde{T}^{\rm tap}_{b}(\mathbf{k})^*}{V N(\mathbf{k})},
\end{equation}
where $V$ is the survey volume and $N(\mathbf{k})$ is a normalization factor derived in Appendix~\ref{sec:windowfunctionderiv}. This is then cylindrically binned by folding $\pm k_z$ into a single $k_\parallel$ coordinate and averaging over rings of constant $k_\perp \equiv (k_x^2 + k_y^2)^{1/2}$ to form $\hat{P}_{ab}(k_\perp, k_\parallel)$. In some (but not all) of the scenarios that we examine in Sections~\ref{sec:basicforecast} and \ref{sec:designer}, this is the stage at which an extra foreground mitigation step is included. This step is the excision of the \tcm foreground wedge \citep{2012ApJ...756..165P,2012ApJ...752..137M,2012ApJ...757..101T,21cmsense1,2014PhRvD..89b3002D,2014PhRvD..90b3018L,2016ApJ...833..242L,2016MNRAS.458.2928C}, which consists of all Fourier modes satisfying
\begin{equation}
\label{eq:wedgeboundary}
    k_\parallel \lesssim k_\perp \frac{H_0 D_c E(z)}{c(1+z)}
\end{equation}
where $E(z) \equiv \sqrt{\Omega_\Lambda + (1+z)^3 \Omega_m}$, $D_c$ is comoving line-of-sight distance to the midpoint of the survey volume, $H_0$ is the Hubble parameter, $\Omega_m$ is the normalized matter density, and $\Omega_\Lambda$ is the normalized dark energy density. Following (possible) foreground wedge excision, we spherically bin our power spectrum in rings of constant $k \equiv (k_\perp^2 + k_\parallel^2)^{1/2}$ to obtain our final estimator of the isotropic power spectrum $\hat{P}_{ab} (k)$.

Our power spectrum estimation methodology was chosen for its simplicity, in order to best showcase trends that are intrinsic to the nature of \tcm-LIM cross-correlations. Our forecasts will therefore be on the conservative side in terms of sensitivity, given that we have neither used optimal frequentist techniques such as an optimal quadratic estimator \citep{Tegmark_1997,1998PhRvD..57.2117B,2011PhRvD..83j3006L,2015PhRvD..91h3514S,2014PhRvD..89b3002D,2014PhRvD..90b3019L}, nor have we implemented a state-of-the-art Bayesian analysis pipeline \citep{2016ApJS..222....3Z,2016MNRAS.462.3069S,2019MNRAS.488.2904S,2019MNRAS.484.4152S,2023MNRAS.520.4443B,2024arXiv240313767B,2024arXiv240319740C}.

\subsection{Window Functions}\label{subsec:window_functions}

Having gone through the imperfections of our simulated instruments and the various choices of our power spectrum analysis pipeline, our estimated power spectrum $\hat{P}_{ab} (k)$ will almost certainly differ in a systematic way (beyond noise fluctuations) from the true power spectrum $P_{ab} (k)$. Our pipeline therefore includes a theory module that computes the joint window function $W_{ab} (k,k^\prime)$ characterizing the joint cross power response of the two instruments. Specifically, the window function\footnote{It is unfortunate that the term \emph{window function} is an overloaded one in the literature, with multiple meanings and definitions. In digital signal processing and radio astronomy papers (e.g., \citealt{2020MNRAS.494.3712L}) it is often used to denote what we define in this paper to be our \emph{tapering function} $B$. In certain portions of the intensity mapping literature (e.g., \citealt{2016ApJ...817..169L,2019PhRvD.100l3522B,2022MNRAS.515.5813P}) it quantifies the instrument's multiplicative response to the true power spectrum, accounting for depressions in power due to a telescope's finite resolution (whether spectral or angular). In our nomenclature, this would correspond to our normalization factor $N(\mathbf{k})$. The normalization factor will only coincide with our more general definition of a window function, defined in Equation~\eqref{eq:windowdef}, when $W_{ab} (k, k^\prime)$ can be approximated as diagonal operator [i.e., proportional to $\delta^D (k-k^\prime)$] and the window functions are assumed to be unnormalized (see Appendix~\ref{sec:windowfunctionderiv} for details).} relates the ensemble average of the estimated power spectrum $\hat{P}_{ab}(k)$ to the true power spectrum $P_{ab}$ via
\begin{equation}
\label{eq:windowdef}
\langle \hat{P}_{ab}(k) \rangle = \int \! dk^\prime W_{ab}(k, k^\prime) P_{ab}(k^\prime).
\end{equation}
Thus, this window function serves as a transfer function that acts on a theoretical input power spectrum to give an output that can be directly compared to the output of our simulation and analysis pipelines. Note that because of the normalization factor included when we estimated our power spectra using Equation~\eqref{eq:PspecEstimator}, integrating the window functions over $k^\prime$ from $-\infty$ to $+\infty$ at fixed $k$ will yield unity by construction. As such, our power spectrum estimates are weighted \emph{averages} (rather than just unnormalized weighted \emph{sums}) over the true power. In Appendix~\ref{sec:windowfunctionderiv} we derive the window functions that are computed by our pipeline and used in the interpretation of our results in Sections~\ref{sec:basicforecast} and \ref{sec:designer}.

\section{Forecasting HERA $\times$ CCAT-prime}
\label{sec:basicforecast}

Using the simulation pipeline presented in the last few sections, we forecast the yield of upcoming cross-correlation measurements from HERA and CCATp. We simulate three redshift bins centered on $z = \{6,7,8\}$ with a width of $\Delta z \sim 0.5$ corresponding a bandwidth of $\sim 10\,\textrm{MHz}$ for \tcm observations and a bandwidth of $13.6\,\textrm{GHz}$ for [CII] observations. In order to capture the non-Gaussian covariance of these upcoming measurements, we perform a set of Monte Carlo simulations over 140 patches of sky as shown in Figure~\ref{fig:FG_patches} to build up a distribution for each measurement. We have three goals here: the first is of course the forecast itself; the second is to use the forecast as a worked example of our pipeline; the final goal is to use our full set of non-Gaussian end-to-end Monte Carlo simulations to extract useful lessons for LIM cross-correlation forecasting, although we defer a description of those results to Section~\ref{sec:sidequestlessons}.

\subsection{Avoiding the Foreground Wedge}
\label{sec:WedgeCutForecast}

\begin{figure*}[ht]
\includegraphics[width=1\textwidth]{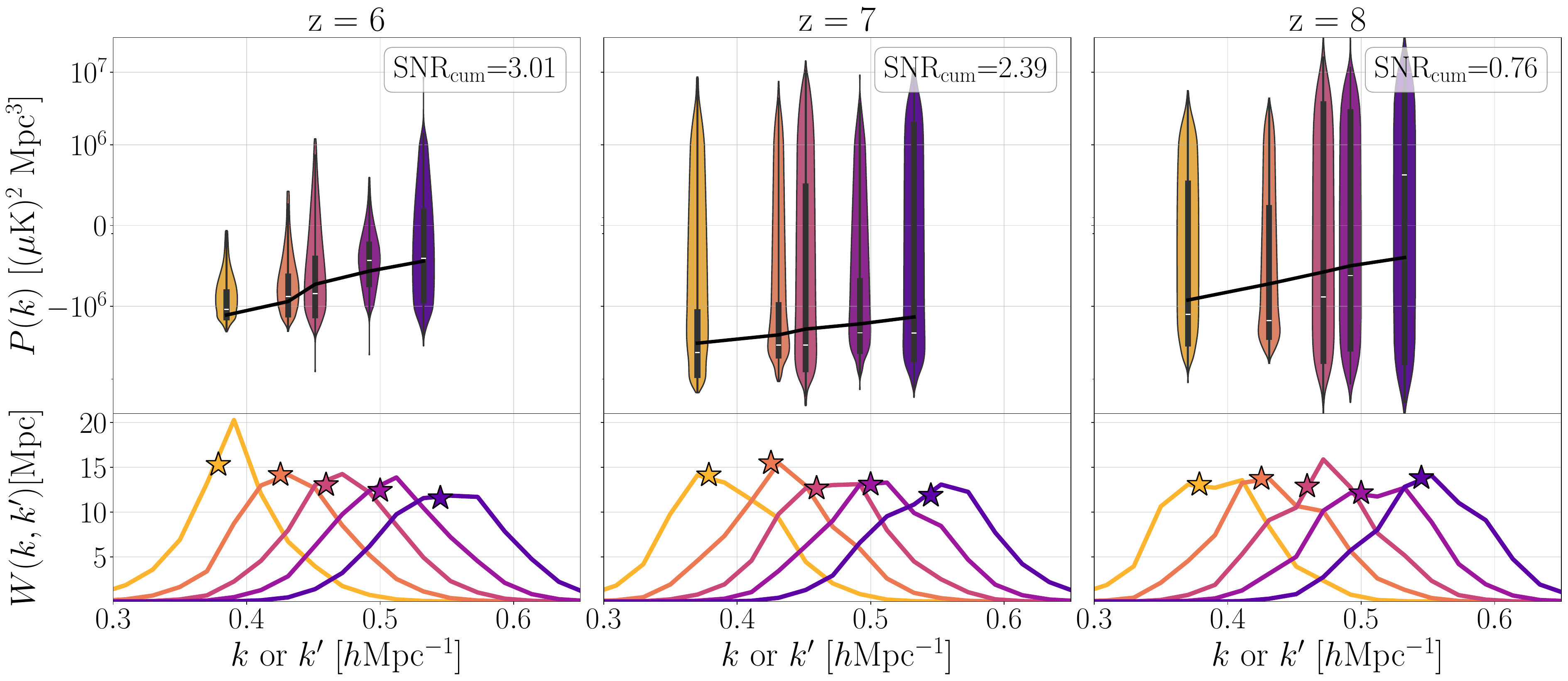}
\caption{Forecasted power spectrum measurement at $z=6$ (left), $z=7$ (middle), and $z=8$ (right) for the HERA $\times$ CCAT-prime scenario described in Section~\ref{sec:WedgeCutForecast}, where foreground-contaminated wedge modes are cut. Top: distributions of power spectrum estimates from our Monte Carlo ensemble, with violins showing the shapes of the distributions, the boxes giving 25th and 75th percentiles, the white bar giving the 50th percentile, and the whiskers showing the full extent of the distributions. Black lines indicate the true inputted power spectra. Note that the vertical axis is a hybrid scale that is linear for $|P(k)| \leq 10^6 \,\mu\textrm{K}\,\textrm{Mpc}^3$ and logarithmic otherwise. The cumulative signal-to-noise ratio, computed using Equation~\eqref{eq:SNRcumformula}, is also indicated. The $z=6$ sensitivity is limited by thermal noise, whereas the $z=7$ and $z=8$ measurements are more severely limited by a CO (2-1) interloper. Bottom: $W(k, k^\prime)$ as a function of $k^\prime$ for each power spectrum estimate in matching colors. Stars indicate the $k$ bin center used in power spectrum estimation. The non-trivial shapes of the $P(k)$ distributions and the window functions showcase the importance of forecasting via end-to-end simulations. The limited Fourier overlap between HERA and CCAT-prime results in a limited cumulative SNR for these measurements.}
\label{fig:violin_HERA_CCAT}
\end{figure*} 

Given that we are forecasting measurements from HERA, we first examine a scenario that uses HERA's current foreground avoidance strategy in which modes that lie within the foreground wedge given by Equation~\eqref{eq:wedgeboundary} are discarded when forming a power spectrum \citep{HERA_H1C_2021,2023ApJ...945..124H}. With our particular instrumental, observational, and data analysis assumptions here, this amounts to computing the cross-power spectrum for $k_{\parallel} > 0.29 h\rm{Mpc}^{-1}$ and $k_\perp < 0.29 h\rm{Mpc}^{-1}$ in 5 $k$-bins. Apart from the wedge cut, we did not perform any other foreground mitigation on either the \tcm or [CII] observations. This scenario is a rather aggressive one that succeeds in avoiding a complete domination of the cross-power spectra by foregrounds, and will serve as a contrasting case when we adjudicate whether a cross-correlation on its own is capable of correlating away foregrounds, as argued in previous work.

In Figure~\ref{fig:violin_HERA_CCAT} we show the cross-power measurements in each of the redshift bins. Our Monte Carlo ensemble enables us to understand the full distribution of each power measurement, which we portray using violin plots. The box and whiskers of each point represent the 25th and 75th percentiles and the extrema of the distributions respectively, and the short white bar within each box is the median value. The solid black line shows the expected truth obtained by evaluating Equation~\eqref{eq:windowdef}. Also shown are the corresponding window functions for each measured power spectrum point. Since these quantify the wavenumber range that is actually probed by each point, we center each violin at the $50$th percentile of $W(k,k^\prime)$. Each window function curve corresponds to $W(k,k^\prime)$ for a fixed value of $k$ (given by the star symbol) plotted as a function of $k^\prime$.

Several generic features are immediately apparent from Figure~\ref{fig:violin_HERA_CCAT}. Broadly speaking, we see from the window functions that when evaluating our power spectrum estimator $\hat{P}_{ab}$ at some target $k$ value (star points), one is indeed roughly probing true power at the correct wavenumber. However, the window functions are rather broad, which is unsurprising for experiments with non-trivial spectral responses \citep{2014PhRvD..90b3018L}. Thus, it would be incorrect to assume that the power is being entirely sourced from the target $k$. In addition, it is not uncommon for there to be some skew to $W(k,k^\prime)$ as function of $k^\prime$, where the stars coincide with neither the peak nor the median. This highlights the importance of computing window functions. Another crucial feature is the presence of visibly non-Gaussian distributions in the violins (although we caution the reader that part of the visual distortion is due to the hybrid vertical scale that transitions from linear to logarithmic at $P(k) = \pm 10^6\,\mu\textrm{K}^2\,\textrm{Mpc}^3$).

The most salient feature of the actual forecasting in Figure~\ref{fig:violin_HERA_CCAT}
is the high variance of the simulated measurements, especially for $z = 7$ and $z = 8$. This can be (crudely) captured by computing a total signal-to-noise ratio (SNR) that is cumulative accross the entire spectrum and is given by
\begin{equation}
\label{eq:SNRcumformula}
    {\rm SNR}_{\rm cum} = \sqrt{\langle \mathbf{\hat{P}} \rangle^t \boldsymbol{\Sigma}^{-1} \langle \mathbf{\hat{P}} \rangle },
\end{equation}
where $\mathbf{\hat{P}}$ is a vector (in this case with five elements) storing the power spectrum measurements for a particular Monte Carlo realization and $\boldsymbol{\Sigma} \equiv \langle \mathbf{\hat{P}} \mathbf{\hat{P}}^t \rangle - \langle \mathbf{\hat{P}} \rangle \langle \mathbf{\hat{P}}\rangle^t$ is the covariance between power spectrum bins, computed by ensemble averaging over our realizations. The ${\rm SNR}_{\rm cum}$ values are 3.01, 2.38, and 0.75 for the $z = 6$, $7$, and $8$ bins, respectively. In our present scenario, therefore, we have (at best) a marginal detection in the $z = 6$ and $z = 7$ bins.

To understand our lack of a highly significant detection, we can use the flexibility of our simulations to decompose our error bars into their constituent sources of uncertainty. Although the error distributions in Figure~\ref{fig:violin_HERA_CCAT} naturally represent the total uncertainty arising from foreground residuals, cosmic variance, and instrumental noise, we can hold each contribution fixed in order to discern which source of variance poses the biggest threat to our measurements. This allows one to prioritize their mitigation strategies. We find that across all redshift bins, cosmic variance is negligible and that (as expected) the foregrounds and thermal noise are the biggest problem. At $z = 6$, the cross-correlation measurement is noise-limited. While HERA and CCATp are each sensitive instruments on their own, the lack of Fourier overlap (plus the wedge cut) leaves very few $(k_\perp, k_\parallel)$ bins over which to average down the noise (recall Figure~\ref{fig:theory_crosspower}). For this reason, the noise variance is comparably high across the $z = 7$ and $z = 8$ bins; however, in those two higher redshift bins, foregrounds also become a significant source of uncertainty. Although the wedge cut does an excellent job at removing \tcm Galactic synchrotron foregrounds across all redshift bins, a bright CO (2-1) interloper line that is absent at $z \sim 6$ appears for the [CII] surveys in our $z \sim 7$ and $ z\sim 8$ bins (see Figure~\ref{fig:CO_CII_intensity}), causing the detection significance to deteriorate.

\subsection{Working in the Wedge} 
\label{sec:WedgelessForecast}

\begin{figure*}[ht]
\includegraphics[width=1\textwidth]{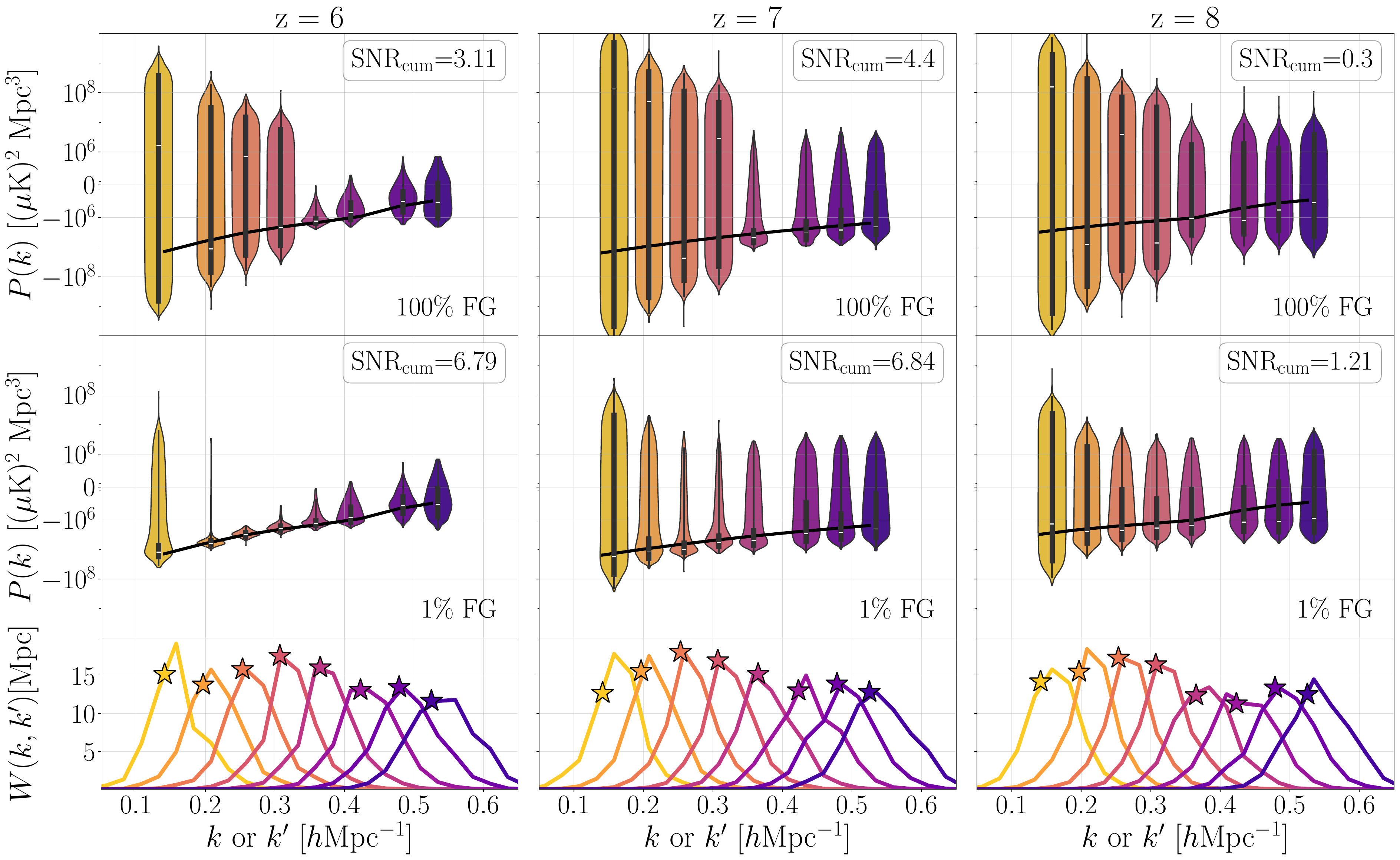}
\caption{Top and bottom: same as Figure~\ref{fig:violin_HERA_CCAT}, but for the scenario described in Section~\ref{sec:WedgelessForecast} where \tcm foreground wedge modes are included. Middle: same as the top, but assuming that \tcm foregrounds have been cleaned to $1\%$ of their original amplitude in the original maps. One sees that cross-correlations alone cannot suppress foregrounds to an acceptable level, but can be powerfully combined with other mitigation techniques to yield statistically significant detections. However, many bins that are important for a detailed characterization of the cross power spectrum remain noise limited.}

\label{fig:NW_violin_HERA_CCAT}
\end{figure*} 

Within the confines of the scenario in Section~\ref{sec:WedgeCutForecast} (limited Fourier overlap plus a wedge cut), one ultimately needs to increase the number of independent measurements in order to increase the signal-to-noise ratio. One method for doing so is to forgo the wedge cut and to push to lower $k$ \citep{2015MNRAS.447.1705P,2016MNRAS.458.2928C}. This is an approach with trade-offs. Including lower $k$ modes has the effect of reducing noise variance, but the residual foreground variance increases, and one is relying more heavily on cross-correlations to reduce what starts out as higher foreground contamination in the data. 

In Figure~\ref{fig:NW_violin_HERA_CCAT} we show the results of including modes in the wedge. In the top panel, where no foreground mitigation has been performed, the large scale modes are foreground variance dominated at all redshifts. At $z=6$ and $z=7$, there is only a slight increase in the cumulative SNR compared to the wedge cut case of the previous section, which highlights the fact that the increased foreground variance outweighs most of the decreased noise variance. In the $z = 8$ bin, the foreground contamination is so extreme that the significance of the detection deteriorates despite the increased number of modes. Relying purely on cross-correlations is simply not enough to suppress foregrounds to a viable level.

In the middle panel of Figure~\ref{fig:NW_violin_HERA_CCAT} we present the results of simulating some \tcm foreground cleaning. To do so we run the same set of Monte Carlo simulations but reduce the \tcm foreground maps to 1\% of their original brightness. This increases the cumulative SNR significantly at $z=6$ and $z=7$ to $6.79$ and $6.84$ respectively due to the decreased foreground variance at low $k$ values. At $z=8$ the cumulative SNR barely breaches unity, which is still not a significant detection.

That a high-significance detection can be made at $z=6$ and $z=7$ having only performed \tcm foreground mitigation is encouraging. In the \tcm auto-correlation literature, much more extreme foreground mitigation is necessary for an auto-spectrum measurement. As a general rule of thumb, the foregrounds are $10^4$ to $10^5$ times brighter than the cosmological signal, necessitating the same $\sim 10^4$ to $10^5$-level foreground suppression for an auto-spectrum measurement \citep{LiuShawReview2020}. Here, we have shown that with 1\% foreground residuals in our \tcm maps, one is able to achieve detections with the assistance of a cross-correlation. In addition, we have performed tests where the CO interlopers were also reduced to the 1\% level and found that it had a negligible effect on the cumulative SNR. This reinforces our intuition that, among the various terms in Equation~\eqref{eq:var_Pcross}, the most problematic are the ones that involve \tcm foregrounds.

\section{A designer's guide for the future}
\label{sec:designer}

\begin{figure*}[ht]
\includegraphics[width=1\textwidth]{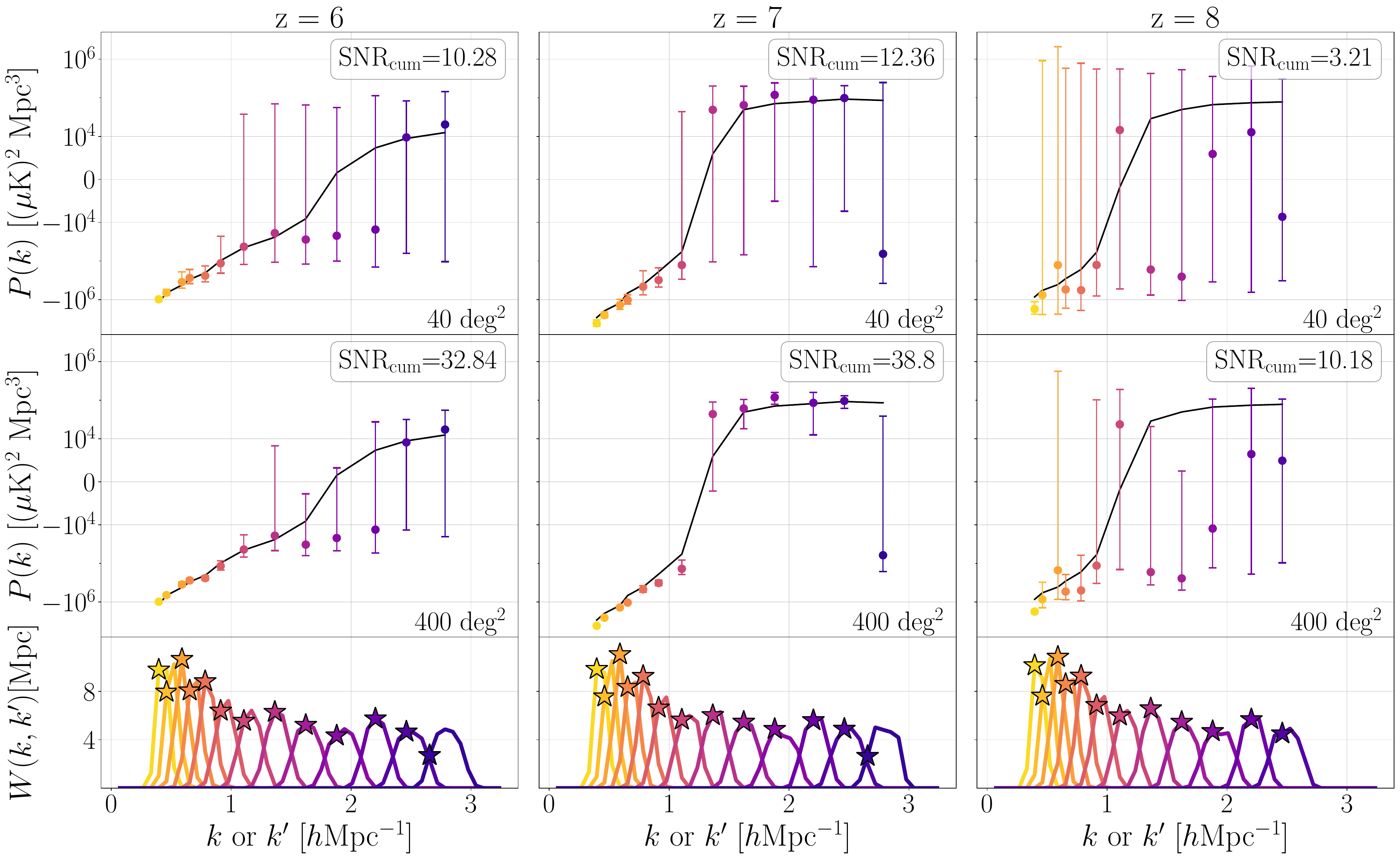}
\caption{Top and bottom: same as Figure~\ref{fig:violin_HERA_CCAT}, but for the futuristic scenario described in Section~\ref{sec:designer} with a wedge cut and where the survey area is 40 deg$^2$ and the spectral resolution of the [CII] is increased. Middle: same as the top, but survey area is increased to 400 deg$^2$. With the increased overlap both in Fourier space (due to increased spectral resolution) and in configuration space (due to increased survey area), the detection significance rises markedly compared to current-generation surveys.}
\label{fig:FW_violin}
\end{figure*} 

\begin{figure*}[ht] 
\centering
\includegraphics[width=0.95\textwidth]{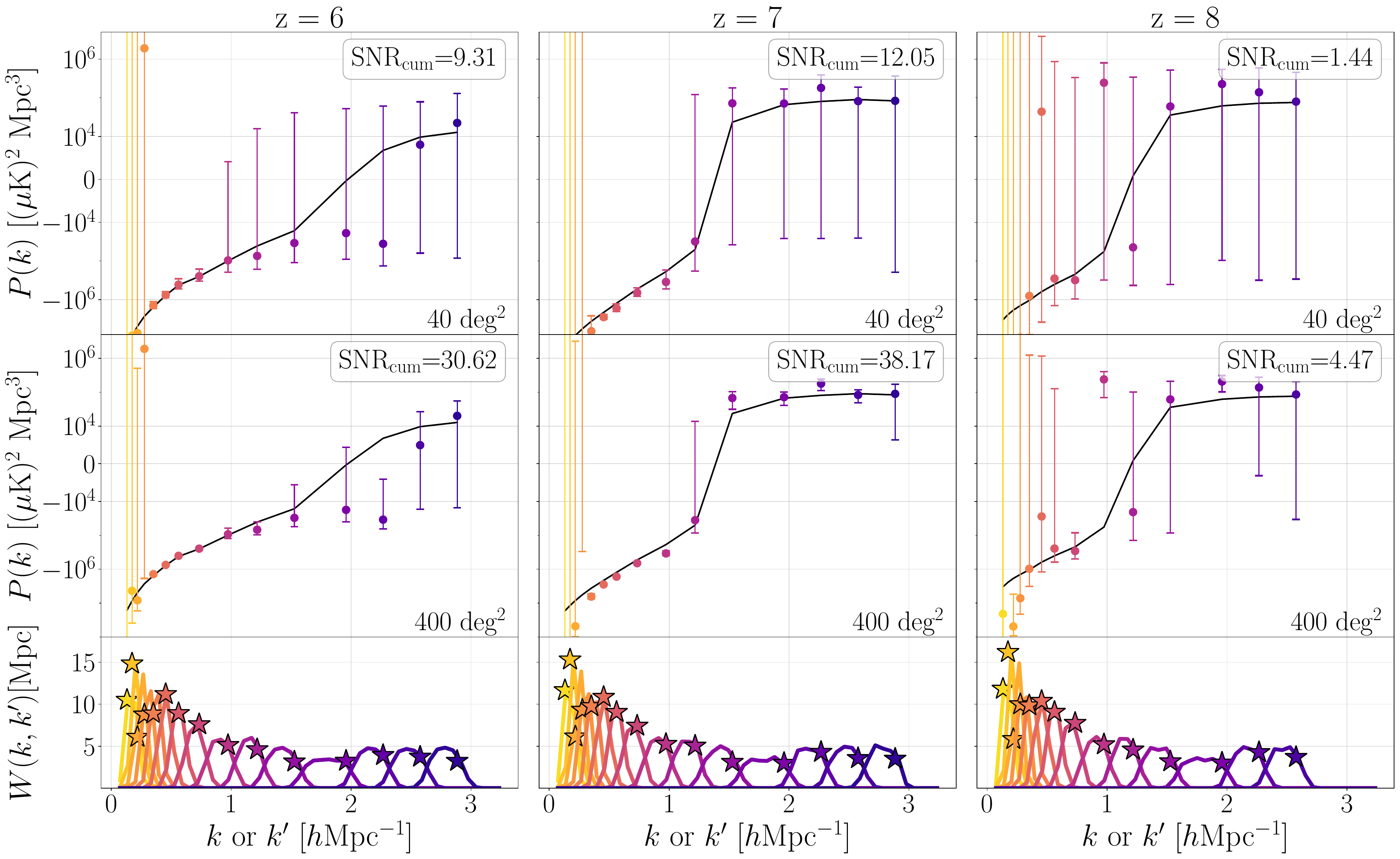}
\caption{Top and bottom: same as Figure~\ref{fig:NW_violin_HERA_CCAT}, but for the futuristic scenario described in Section~\ref{sec:designer} with increased survey area to 40 deg$^2$ and a [CII] instrument with greater spectral resolution, but no \tcm foreground mitigation nor avoidance of the wedge. Middle: same as the top, but survey area is increased to 400 deg$^2$.}
\label{fig:FNW_1.0_violin}
\includegraphics[width=0.95\textwidth]{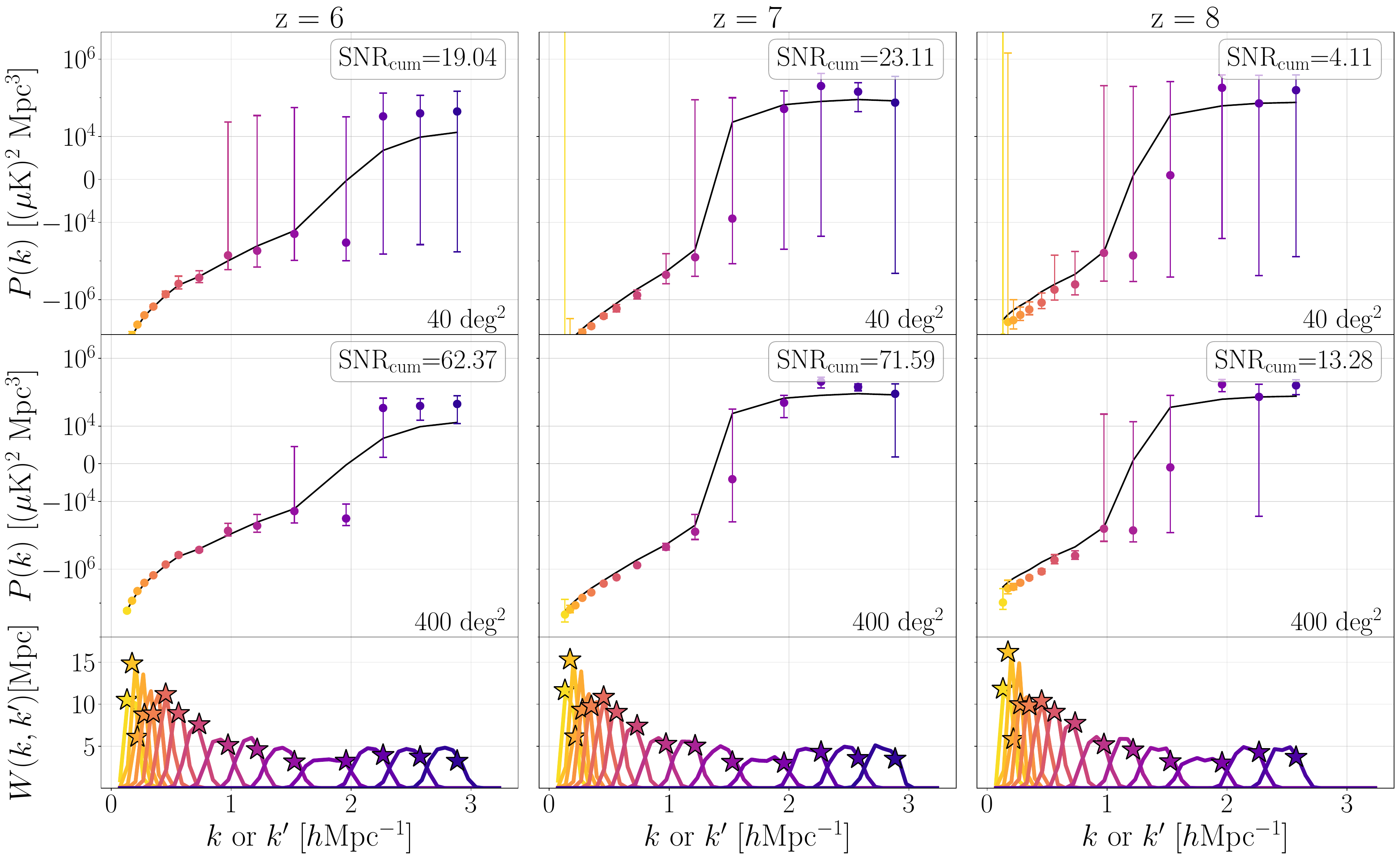}
\caption{Same as Figure~\ref{fig:FNW_1.0_violin}, but assuming that foregrounds in the \tcm maps have been reduced to $1\%$ of their original level. Compared to Figures~\ref{fig:FW_violin} and \ref{fig:FNW_1.0_violin}, the extra foreground cleaning allows individual bins of the cross power spectrum to be detected to high significance, permitting a detailed characterization of the spectrum and inference of parameters such as the crossover scale from negative to positive correlation.}
\label{fig:FW_0.01_violin}
\end{figure*} 

Figure~\ref{fig:NW_violin_HERA_CCAT} suggests that pushing to lower $k$ has some utility when  combined with some \tcm foreground cleaning. A future-looking complement that we now examine is to go to wider survey areas and to increase the number of independent Fourier modes by going to higher $k$. The latter has several advantages. First, from the perspective of increasing cumulative SNR, going to higher $k$ simply adds more modes that can be used to beat down the variance. Second, accessing higher $k_\parallel$ provides access to new independent modes that are less foreground-contaminated. Finally, going to a broader range of $k$ increases the possibility of catching the (as-yet unknown) crossover scale when the cross-correlation power spectrum transitions from being negative to positive. Probing these scales would give us access to information about the morphology of ionization field and allow us to constrain a wider range of reionization scenarios.

In this section we forecast a futuristic \tcm-[CII] cross-correlation measurement where HERA and CCATp have increased Fourier overlap in the $k_\parallel$ direction up to $k_\parallel \approx 3\,h\rm{Mpc}^{-1}$. This requires increasing the CCATp spectral resolution to $\Delta\nu = 0.5\,\textrm{GHz}$
($R\approx 500$). Apart of this modification, the specifications for each instrument remain the ones listed in Tables~\ref{tab:HERA} and \ref{tab:ccat}. For our more futuristic scenarios, we simulate measurements over $40\,\textrm{deg}^2$ and $400\,\textrm{deg}^2$. Rather than simulating a contiguous survey area, we continue to simulate the patches shown in Figure~\ref{fig:FG_patches} but average together the power spectrum results from multiple patches to accumulate the larger survey areas. Although this represents an incoherent average of power spectra (rather than a coherent increase in contiguous survey area prior to forming power spectra), our calculations should serve as a reasonable proxy for the expected sensitivity since we are not seeking to access lower $k_\perp$ modes that are angularly coherent over large parts of the sky. To gather statistics for increased survey areas, we perform bootstrap sampling over the different foreground patches. As in the previous section, we simulate two analyses: one with a wedge cut and one without. 

In Figure~\ref{fig:FW_violin} the cross spectra with the wedge cut are shown. To avoid visual clutter, we forgo the violin plots and instead show conventional error bars, with the expectation that the increased averaging implicit in our new scenarios will serve to somewhat Gaussianize the distributions (see Section~\ref{sec:GaussErrs} for a critical examination of this). One sees that over 40 deg$^2$, precise measurements can be made of the large scale modes at $z = 6$ and $z = 7$. While in principle the increased frequency resolution of the [CII] survey ought to allow one to measure the crossover scale, in practice the high-$k$ bins remain noise limited. A brief examination of Figure~\ref{fig:theory_crosspower} reveals the reason for this. Since our futuristic scenario does not entail adding coverage at higher $k_\perp$ (which would be extremely difficult for \tcm experiments), the new high $k$ bins are formed from relatively small sections of contours of constant $k \equiv \sqrt{k_\perp^2 +k_\parallel^2}$ that are almost horizontal on the $k_\perp$-$k_\parallel$ plane. Thus, relatively few independent modes are added, limiting one's ability to average down the noise. For surveys conducted over $400\,\textrm{deg}^2$ many of the trends remain, although measurements across the crossover scale are now possible in the $z = 7$ bin.

In Figure~\ref{fig:FNW_1.0_violin} the spectra without the wedge cut and without any foreground mitigation are shown. Due to the highly contaminated foreground wedge modes, there is a decrease in the SNR in comparison to when the wedge cut was performed.\footnote{At first glance, this may seem to be a counterintuitive result: one might expect that including extra Fourier modes would only increase the cumulative SNR, even if those modes have an extremely large variance to them. In some ways, this is an artifact of our simple power spectrum estimator and the way we compute $\textrm{SNR}_\textrm{cum}$. Our cumulative SNR is computed in a spherically averaged space, that is, for the power spectrum estimator $\hat{P}_{ab}$ as a function of $k$ rather than $k_\perp$ and $k_\parallel$. Many of our $k$ bins straddle the foreground wedge, with some modes coming from within the wedge and some modes coming from outside the wedge. Now, recall that our simple estimator $\hat{P}_{ab}(k)$ simply averages $\hat{P}_{ab}(k_\perp, k_\parallel)$ in rings of constant $k$ \emph{with uniform weights all modes}. This means that without the excision of the wedge, modes with extremely high foreground variance can otherwise contaminate what would be a $k$ bin with reasonable SNR. A more advanced treatment would be to perform power spectrum estimation using an optimal quadratic estimator. Such an estimator would downweight the data by its total inverse covariance matrix (including the foreground covariance) \emph{before} binning \citep{2014PhRvD..90b3018L}, thereby organically preventing a contaminated $k_\perp$-$k_\parallel$ foreground mode from polluting a $k$ bin \citep{2011PhRvD..83j3006L}. In fact, the Monte Carlo approach that we espouse in this paper can provide precisely the needed covariance information to do this, but for simplicity we leave this investigation to future work.} That being said, many modes can still be measured to high significance especially over the larger 400 deg$^2$ survey area. If one, once again, performs \tcm foreground mitigation with 1\% residuals, the situation drastically improves. In Figure~\ref{fig:FW_0.01_violin} we present the spectra when foreground mitigation is performed. In this case, precision measurements of large scale modes can be made across all redshifts  for a $40\,\textrm{deg}^2$ survey. Over $400\,\textrm{deg}^2$, precision measurements of both large-scale modes and small-scale modes can be made at all redshifts. By designing a CCAT-like instrument with increased spectral resolution, optimizing survey strategies to ensure extensive sky coverage overlap, and coupling these improvements with a modest foreground mitigation strategy, a high-significance detection of the \tcm--[CII] cross spectrum is likely achievable.

\section{Lessons Learned}
\label{sec:sidequestlessons}
In conducting the forecasts of Sections~\ref{sec:basicforecast} and \ref{sec:designer}, our Monte Carlo simulations allow considerable flexibility for exploring forecasting methodology. In this section, we summarize some of the ``lessons learned" along the way, with an eye towards understanding which commonly used forecasting approximations are justifiable and which are not.

\subsection{A $1/\sqrt{N}$ Scaling Works for Foregrounds}

Implicit in the use of cross-correlations for foreground mitigation is the assumption that different patches of the sky have independent foregrounds. In effect, one is using angular averages as an approximation for the ensemble averaging in Equation~\eqref{eq:x-spec full}, and the hope is that by increasing sky coverage, Equation~\eqref{eq:x-corr simplified} becomes a better and better approximation. Crucially, this relies on different parts of the sky having independent foregrounds.

In principle, foregrounds are most certainly not independent from one part of the sky to another on large scales. This is clear from a visual inspection of Figure~\ref{fig:21cm_foregrounds}, or from the existence of large-scale power in previous foreground observations \citep{WMAP_FG_2009,WMAP_FG_2011,Planck_FG_2015}. However, in practice it may be a sufficiently good approximation for foreground \emph{fluctuations}---as quantified by the foreground contribution in a particular $\mathbf{k}$ mode---to be treated as independent when sourced from different parts of a survey.
\begin{figure}[h]
\includegraphics[width=0.5\textwidth]{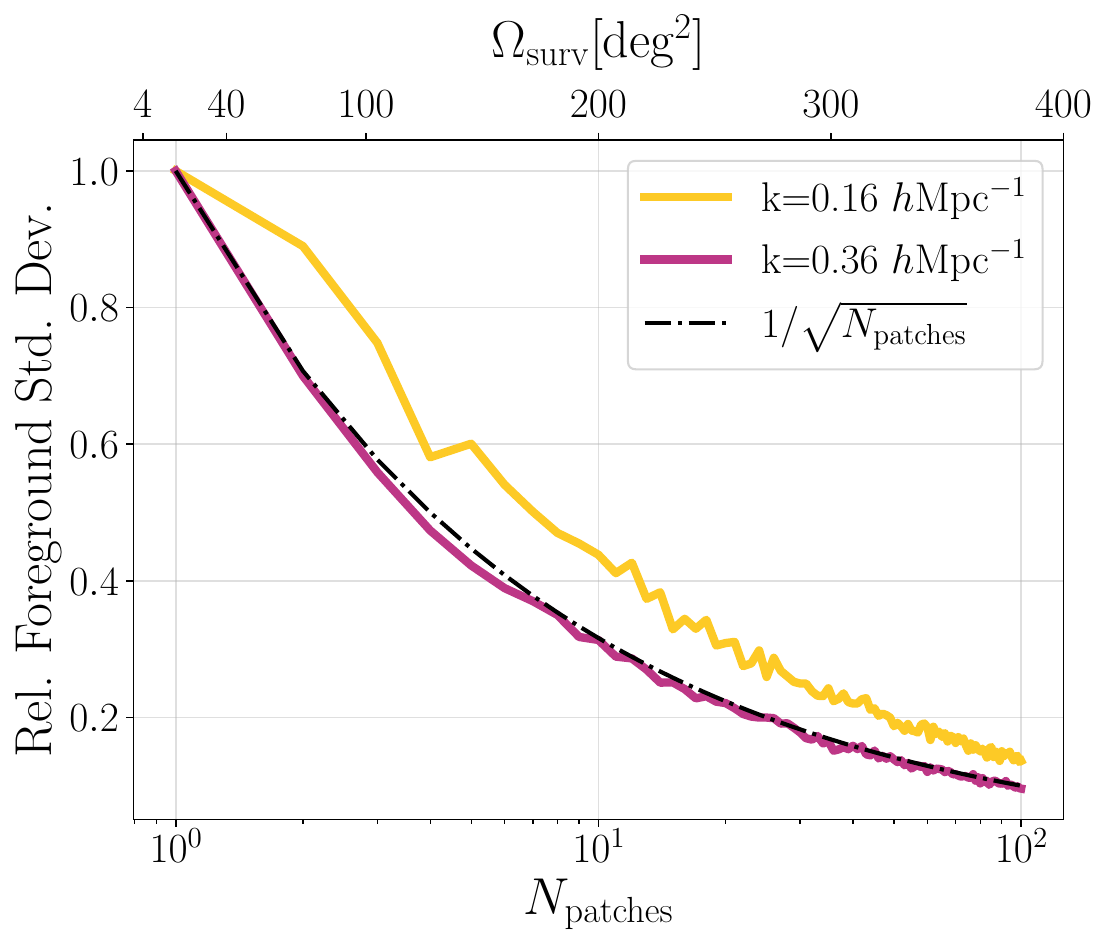}
\caption{Relative foreground standard deviation as a function of number of simulation patches averaged together (equivalently survey area). In yellow the foreground standard deviation for the $k=0.16$ $h$Mpc$^{-1}$ is plotted, and in pink the same quantity for $k=0.36$ $h$Mpc$^{-1}$ is plotted. As a reference, the expected $1/\sqrt{N_{\rm patches}}$ relation for $N_{\rm patches}$ independent sample is shown in dashed black. One sees that the foreground residual power averages down with increasing sky area.}
\label{fig:boostrap}
\end{figure}

With our Monte Carlo samples including different draws from different parts of the sky (recall Figure~\ref{fig:FG_patches}), our pipeline is equipped to test (rather than assume) the independence of different foreground patches. We examine the averaging down of foregrounds in two different power spectrum bins: one centered on $k = 0.16\,h\textrm{Mpc}^{-1}$ (intended to be at low enough $k$ to be representative of a mode where strong foreground contamination resides) and one centered on $k = 0.36\,h\textrm{Mpc}^{-1}$ (at higher $k$ where there is weaker but still non-negligible foregrounds from leakage effects). For each mode, we vary the number of $6\times 6\,\textrm{deg}^2$ patches that are used to estimate the power spectrum. The patches are combined in an incoherent manner, such that power spectra are computed for each individual patch and then averaged together.\footnote{In principle, this will result in a lower sensitivity than a setup where a contiguous volume is analyzed coherently. As an extreme example of this, one's sensitivity to angular modes larger than the extent of an individual patch is formally zero when the patches are analyzed separately but non-zero when analyzed together. In practice, our forecasts do not deal with sufficiently small $k_\perp$ values for this to be a concern.} This yields a single power spectrum estimate with suppressed (but still existent) foreground residuals that have not entirely cross-correlated away. To compute the increase in foreground variance that this implies, we perform a bootstrap sampling over our random selection of patches in Figure~\ref{fig:FG_patches}.

In Figure~\ref{fig:boostrap} we show the results of such an experiment, with the variance of our bootstrapped samples normalized to the variance for a single $6\times 6\,\textrm{deg}^2$ patch. The dark purple line (showing the behaviour of the $k = 0.36\,h\textrm{Mpc}^{-1}$ bin) clearly follows a $1/\sqrt{N}$ form, implying that it is reasonable to assume that the foregrounds average down independently. The lighter yellow line (for $k = 0.16\,h\textrm{Mpc}^{-1}$) deviates slightly from $1/\sqrt{N}$ but still averages down quite well as more patches are included. One may thus conclude that although the gold standard for forecasting would be to simulate precisely the planned sky coverage for one's survey, for rough back-of-the-envelope estimates it is likely acceptable to treat different foreground patches as independent in a cross-correlation experiment. This validates, for example, the approach used in \citet{roy2023crosscorrelation} of simulating smaller volumes and then appropriately scaling the resulting SNRs to match proposed surveys, even if foregrounds are involved.

\subsection{Some Gaussianization Of Errors Distribution Occurs But With Residual Non-Trivial Behaviour In The Tails}
\label{sec:GaussErrs}

A common assumption in many forecasting efforts is that error bars can be treated as Gaussian (although see \citealt{2023MNRAS.521.5191W} for a critical examination in the context of \tcm auto power spectrum measurements). Again, our pipeline is equipped to test this assumption, since we include the non-Gaussian effects of foregrounds and furthermore propagate its interactions with other sources of uncertainty in our measurements. One should expect that with sufficient binning of power spectra that the error distributions would eventually Gaussianize via the Central Limit Theorem (CLT). However, it is crucial to understand whether the probability distributions of foregrounds (in various Fourier modes) are sufficiently well-behaved to satisfy the assumptions of the CLT. This is not \emph{a priori} true for three reasons. First, the probability distributions of foregrounds may have heavy tails to be well-behaved enough for the CLT. Second, the precise mix of foreground, noise, and cosmological power within a spherical Fourier $k$ shell will depend on the direction of $\mathbf{k}$ (indeed, this is the central fact leveraged by wedge-excision foreground mitigation schemes). Third, as we will see in Section~\ref{tab:SNR_exploration}, power spectrum estimates in different parts of Fourier space tend to be correlated with each other. These last two caveats mean that the act of binning from $P(\mathbf{k})$ to $P(k)$ is tantamount to the summation of \emph{non-independent} and \emph{non-identically} distributed random variables---a stark contrast to the independent and identically distributed draws assumed by the CLT.

Although there exist some generalizations of the CLT that can be used for an analytic treatment \citep{billingsley2012wiley,2023MNRAS.521.5191W}, here it is simpler to test for Gaussianization (or lack thereof) empirically using our simulations. To do so, we use the data from our futuristic case with no wedge cut (Figure~\ref{fig:FNW_1.0_violin}), but with the original survey area\footnote{Our motivation for using a smaller survey area was to allow for a larger number of samples without the added layer of complication that comes with bootstrapping, which recall from Section~\ref{sec:designer} is how we perform our forecasts for large survey areas.} of $2 \times 2\,\textrm{deg}^2$. We use our suite of realizations to study the distribution of the cross spectrum power in individual voxels in three-dimensional Fourier space (i.e., the power spectrum estimates prior to any binning). These Fourier voxels are selected such that they would be binned into the same bin in a spherical binning. The distributions of individual voxel values are then compared to the distribution of the final spherically averaged power spectrum. The hope is that while the data in individual Fourier voxels may not be Gaussian distributed, the distribution of the binned value Gaussianizes.

\begin{figure}[]
\includegraphics[width=0.48\textwidth]{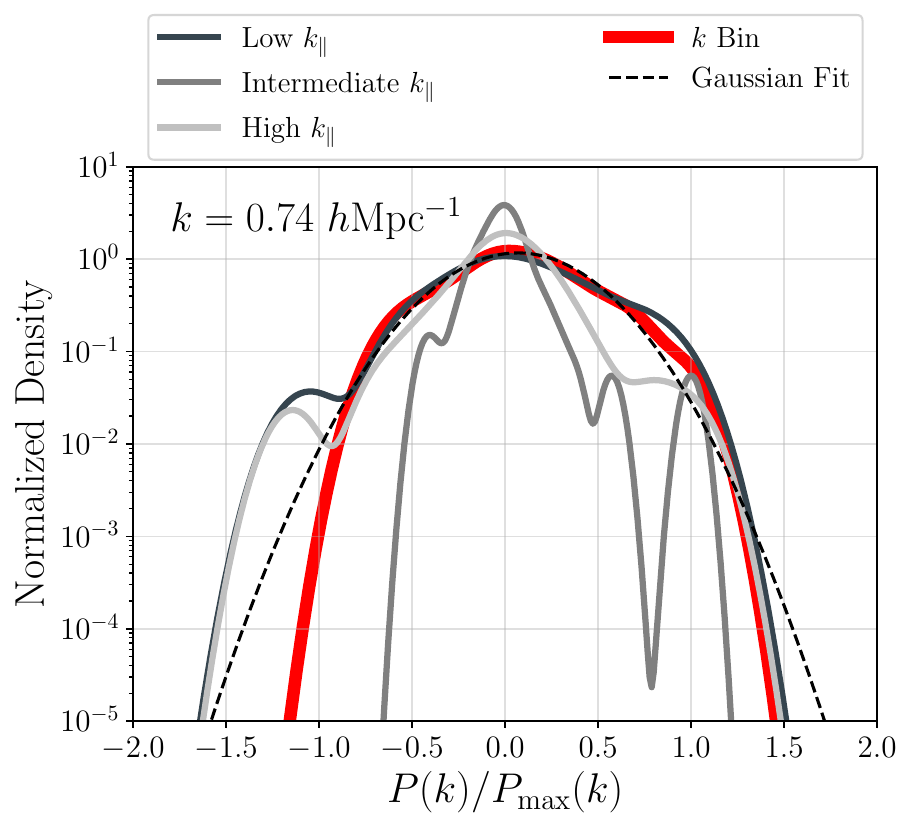}
\caption{The normalized probability density of a selection of power spectrum voxels (various shades of grey) and the resultant power spectrum bin (red). As a reference, the Gaussian fit to the red curve is plotted in dashed black. Binning is seen to Gaussianize power spectrum distributions to some extent, but there are residual discrepancies in the tails.}
\label{fig:morphing_hist}
\end{figure}

In Figure~\ref{fig:morphing_hist} we plot the various distributions.\footnote{For visualization purposes, in both Figure~\ref{fig:morphing_hist} and \ref{fig:errors_histogram} we use a kernel density estimator (with bandwidth parameter set by Scott's rule; \citealt{scott2015multivariate}) rather than directly plotting histograms.} Shown in different shades of grey are the distributions of cross power at Fourier voxels with different $\mathbf{k}$ but roughly identical $k$ that will eventually be averaged into the same $k \approx 0.74 h$ Mpc$^{-1}$ bin. We choose to plot the distributions of three voxels that are representative of the type of modes in the bin: a mode with a low $k_\parallel$ component that is consistently foreground dominated across all realizations, a mode with a high $k_{\parallel}$ component that is consistently much cleaner across all realizations, and finally an intermediate mode. These distributions have extremely different widths to them, since the foregrounds are so much brighter than instrumental noise. Thus, to highlight the shapes of the distributions (rather than their widths), we scale each distribution by its maximum power value $P_\textrm{max}$. Plotted in red is the distribution of the final binned power. The distributions of the individual Fourier voxels are clearly non-Gaussian, exhibiting a multitude of peaks and valleys. The distribution of the resultant bin shown in red is better behaved and is well fit by a Gaussian distribution (dashed black curve) near the peak; however, it fails to Gaussianize in the tails. We thus conclude that a Gaussian approximation is a reasonable treatment out to several standard deviations, but the behaviour beyond that should be treated with care.

\subsection{Cross-Terms are Important, but Separate Simulations are Acceptable}\label{sec:cross_terms_important}

\begin{figure*}[t]
\includegraphics[width=1\textwidth]{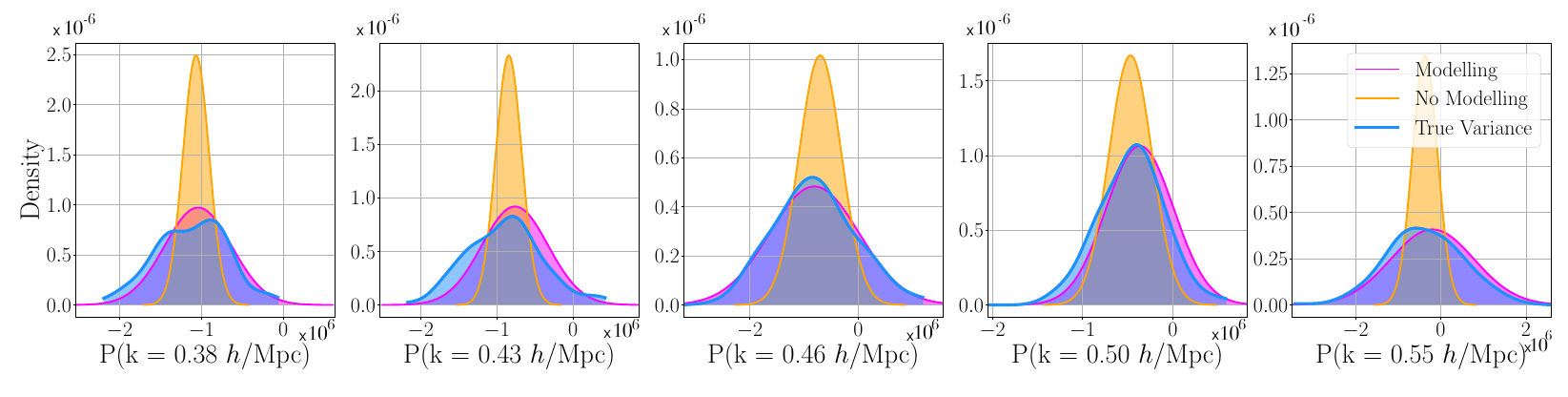}
\caption{From left to right, the probability distribution of the measured cross-power for increasing wavenumber $k$. In blue are the distributions obtained from full MC simulations, in pink are Gaussianized approximations in which each model component is included but not co-varied, and in orange each source of variance is simply computed independently. As more approximations are made, the distributions deviate further from the blue distributions. }
\label{fig:errors_histogram}
\end{figure*}

In this paper, we have stressed the importance of simultaneously varying multiple components of a measurement (i.e., the cosmological signal, foregrounds, and noise) in order to capture the full probability distributions of final power spectrum band powers. The non-Gaussianity of the foregrounds (or in principle the cosmological signal) means that the analogous expression to Equation~\eqref{eq:var_Pcross} for higher-order moments includes terms with non-trivial products between sky components. That the components are statistically independent allows, say, a six-point function with four copies of the foregrounds and two copies of the signal to be reduced a four-point function of foregrounds multiplied by the cosmological power spectrum. That said, taking advantage of such  simplifications still require evaluating higher order moments of each component, which means that in practice it may be easier to simply simulate all components together.

For an approximate treatment, however, one can use the fact (established above) that the non-Gaussianities in the final errors are small in many cases of interest. This suggests that simulating the components separately and then applying Equation~\eqref{eq:var_Pcross} may be sufficient. Figure~\ref{fig:errors_histogram} shows that this is indeed the case for non-wedge modes. In blue we plot the full probability distribution of each mode for the measured HERA and CCATp cross-spectrum from the Monte Carlo simulations used to make Figure~\ref{fig:violin_HERA_CCAT}. These distributions are non-Gaussian, especially at low $k$ values where foregrounds are important. In pink, we plot the error distributions obtained by individually varying the foreground, noise, and cosmic variance components in our simulation. While varying a given component, the others are still modeled and fixed to a single realization. We then compute the standard deviation of each such distribution (effectively making a Gaussian approximation) and add these individual error components in quadrature to obtain the total standard deviation. This standard deviation is inserted into a Gaussian form for the probability distribution. While these Gaussianized distributions obviously do not capture the full shape of the distributions, they are a relatively faithful description of its width and therefore the variance. Finally, we again vary each component separately but in doing so, do not include a realization of any other components. In other words the cosmic variance, noise variance, and foreground variance are computed separately and then added in quadrature. This is equivalent to computing only the first four terms of Equation~\ref{eq:var_Pcross}. The resultant Gaussian distribution is plotted in orange. In this case, since no cross terms have been modeled, the variance is heavily underestimated. Neither the width nor the shape of the distribution captures the variance of the data. 

In the first few columns of Table \ref{tab:SNR_exploration}, we summarize the results of this exploration for current-generation surveys. Approximate methods tend to have the effect of overestimating SNR. Across all redshifts, not modeling the full variance (``No Modeling" column) leads to a overestimation of the cumulative SNR. When no joint modelling is performed, the cumulative SNR is overestimated by over 5.41$\sigma$ at $z = 6$, by 0.40$\sigma$ at $z = 7$, and by 0.8$\sigma$ at $z = 8$. When some modeling of the other simulation components is included (``Modeling" column), the cumulative SNR is a much more faithful representation of the true variance, although one loses  the full shape of the distribution. The results from simultaneously varying all components and computing the variance directly form the full non-Gaussian distributions in Figure~\ref{fig:errors_histogram} are shown in the column labeled ``Var". One sees that the SNR decreases further. In the following section, we will see yet another effect when we explore the importance of not only considering the non-Gaussianities but also error covariances between bins.


\begin{table}[ht]
\centering
\setlength{\tabcolsep}{11pt} 
\renewcommand{\arraystretch}{1} 
\begin{tabular}{@{}ccccc@{}}
\toprule
$z$ & No Modeling & Modeling  & Var. & Cov.\\ \midrule
6 & 8.85 & 4.36 & 3.44 & \textbf{3.01} \\
7 & 2.09 & 1.78 & 1.69 & \textbf{2.38} \\
8 & 0.75 & 1.70 & 1.43 & \textbf{0.81} \\ \bottomrule
\end{tabular}
\caption{Cumulative SNR for each redshift bin computed using various approximations. The column labelled ``Cov." makes use of the full covariance matrix and Equation~\ref{eq:SNRcumformula} to compute the SNR. The neighbouring column labelled ``Var." only takes into account the diagonal elements of the covariance matrix (i.e. the variance). The remaining two columns implement further approximations as described in Section~\ref{sec:cross_terms_important}.}
\label{tab:SNR_exploration}
\end{table}

\subsection{Error Correlations Between Bins are Non-Trivial}

\begin{figure*}[t]
\includegraphics[width=1\textwidth]{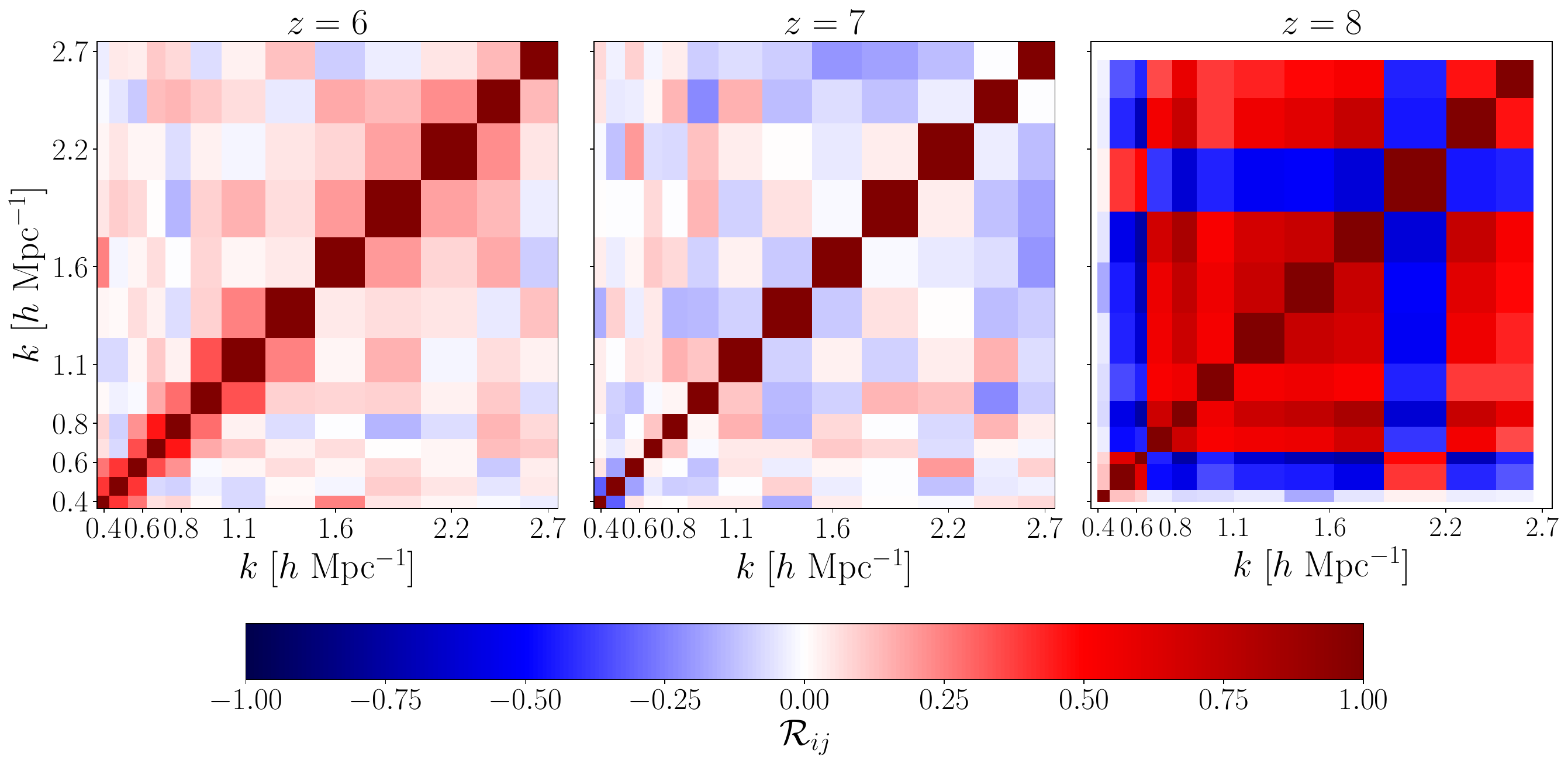}
\caption{Dimensionless correlation matrices $\mathcal{R}$, defined in Equation~\eqref{eq:curlyRdef}, for the futuristic HERA $\times$ CCAT-prime forecasts shown in Figure~\ref{fig:FW_violin}. Substantial error correlations exist between all $k$ bins, necessitating the incorporation of full covariance information in evaluating signal-to-noise or in propagating power spectrum estimates to downstream parameter inferences.}
\label{fig:covariances}
\end{figure*} 

A frequently used rule-of-thumb in performing power spectrum sensitivity forecasts is to assume that the errors in different Fourier bins are independent as long as their wavenumber spacing $\Delta k$ is greater than $\Delta k \sim 2\pi / L$, where $L$ is a characteristic size of a survey \citep{1998ApJ...499..555T}. With our suite of simulations, we can put this assumption to the test.

In order to highlight the interdependence (or lack thereof) of different Fourier bins, we compute the dimensionless correlation matrix $\mathcal{R}$, given by
\begin{equation}
\label{eq:curlyRdef}
    \mathcal{R}_{ij} \equiv \frac{\boldsymbol{\Sigma}_{ij}}{\sqrt{\boldsymbol{\Sigma}_{ii} \boldsymbol{\Sigma}_{jj}}},
\end{equation}
rather than the covariance matrix $\boldsymbol{\Sigma}$. The results are shown in  Figure~\ref{fig:covariances}, using the forecasts of Section~\ref{sec:designer} with a wedge cut as an example. One sees that substantial, non-negligible correlations exist between different bins. This is not particularly surprising, since the assumption of independent bins can be violated by complicated instrumental responses \citep{2016ApJ...833..242L}, or analysis steps such as apodization and the downweighting or projecting out of systematics (such as foregrounds; \citealt{2014PhRvD..90b3018L}). \citet{JoseDeInterloping} have also shown that residual interloper contaminants will naturally result in error correlations. Continuum foregrounds are also likely to induce correlation between bins (particularly in the regime where $k \approx k_\parallel$), since a key source of variance comes from the leakage of residual foreground power from low $k$ (where the foregrounds intrinsically reside) to high $k$. With all these complicated and interconnected effects, it is no surprise that the covariance structure seen in Figure~\ref{fig:covariances} is non-trivial, which speaks to the necessity of multi-component end-to-end simulations. For example, these error correlations are important to incorporate especially in low-SNR regimes. In the column labeled ``Cov." of Table~\ref{tab:SNR_exploration}, we provide the SNRs computed using the full error covariances for the scenario described in Section~\ref{sec:WedgeCutForecast} (i.e. HERA $\times$ CCATp with a wedge cut). At all redshifts, these values differ from the previously discussed ones that assumed no error covariance (column labeled ``Var.").

What we have seen here is that modeling assumptions do matter in determining the cumulative SNR. Although we expect this to be true even for advanced futuristic experiments, our illustrative example here was computed for current-generation experiments because the metric of cumulative SNR is most appropriate when one is chasing an initial detection. Once one is \emph{characterizing} a high-significance measurement, the details of precisely which parts of a spectrum are measured begin to matter, and a cumulative SNR becomes too blunt a metric. For example, while it is clear that a high-precision \tcm $\times$ [CII] measurement at \emph{precisely} the crossover scale between the negative and positive correlation regimes would be extremely useful, Equation~\eqref{eq:SNRcumformula} would ascribe precisely zero value to such a measurement! Said differently, SNR studies are necessary---but not sufficient---for understanding instrument performance, and in the following section we explore the effects of  error covariances on a simple toy example of parameter estimation.

\subsection{Window Functions Should Be Computed and Propagated Downstream for Parameter Estimation}\label{sec:inference}

Given the extra effort required to compute window functions as well as full error covariances, one may wonder how crucial this is when it comes to parameter estimation. In order to demonstrate the importance of modelling these quantities, we present a toy model whose parameters we seek to constrain and showcase how those constraints change in the presence of various error correlations and window functions. 

Suppose one seeks to constrain the crossover scale $k_0$ at which the cross spectrum between [CII] and \tcm transitions from negative to positive, as discussed in Section~\ref{sec:Motivation_PipelineOverview} and illustrated in Figure~\ref{fig:Fields+Spectra}. As a simplified model, we imagine taking the two power spectrum points $\mathbf{P}_{\pm} \equiv (P_-, P_+)$ on either side and fitting a straight line model $P^{\rm mod}(k)$ between them to find the crossover, such that
\begin{equation}
\label{eq:linearmod}
P^{\rm mod}(k) \approx m (k-k_0),
\end{equation}
where $m$ is the slope of the line. Finding the crossover scale is then tantamount to computing the posterior distribution for $k_0$ given the measured power spectrum values, the error covariance between them, and their associated window functions. This posterior distribution $p(k_0| \mathbf{P}_\pm, \boldsymbol{\Sigma}, W)$ is given by
\begin{equation}
    p(k_0| \mathbf{P}_\pm, \boldsymbol{\Sigma}, W) = \int \! dm \, p(k_0, m| \mathbf{P}_\pm, \boldsymbol{\Sigma}, W),
\end{equation}
where $W$ contains the window functions for the two bandpowers, and $p(k_0, m| \mathbf{P}_\pm, \boldsymbol{\Sigma}, W)$ is the joint posterior for the slope and the crossover scale. Bayes' theorem \citep{bayes1763lii} enables this to be written as
\begin{equation}
\label{eq:k0mposterior}
    p(k_0, m| \mathbf{P}_\pm, \boldsymbol{\Sigma}, W) \propto p(\mathbf{P}_\pm| k_0, m, \boldsymbol{\Sigma}, W) p(k_0, m| \boldsymbol{\Sigma}, W),
\end{equation}
where $p(k_0, m| \boldsymbol{\Sigma}, W)$ is the prior and $p(\mathbf{P}_\pm| k_0, m, \boldsymbol{\Sigma}, W)$ is the likelihood. Given our previous demonstrations that the binned power spectra have Gaussianized to some degree (at least away from the low-level tails of the distribution), the logarithm of the likelihood can be written as\footnote{In Equation~\eqref{eq:k0mposterior}, we have omitted the normalization term of the Gaussian likelihood. In principle, cosmological information is present in this term, since $\boldsymbol{\Sigma}$ is the total covariance---including contributions from the cosmological signal (i.e., cosmic variance). In practice, for most experiments the constraining power from this piece of the likelihood is negligible \citep{1997PhRvD..55.5895T}, so we neglect it.}
\begin{equation}
\label{eq:loglike}
    \ln p(\mathbf{P}_\pm| k_0, m, \boldsymbol{\Sigma}, W) \propto  -\frac{1}{2} (\mathbf{P}_\pm - \mathbf{P}_\pm^{\rm mod})^t \boldsymbol{\Sigma}^{-1} (\mathbf{P}_\pm - \mathbf{P}_\pm^{\rm mod}),
\end{equation}
where $\mathbf{P}_\pm^{\rm mod} \equiv (P_-^{\rm mod}, P_+^{\rm mod})$ is Equation~\eqref{eq:linearmod}, i.e., our model, evaluated at the $k_-$ and $k_+$, the wavenumbers of the bandpowers just shy of and just beyond the crossover scale, respectively. These are given by
\begin{eqnarray}
    P_\pm^{\rm mod} &=& \int \! dk \, W(k_\pm, k) P^{\rm mod}(k) \nonumber \\
    &=& m \left[ \int \! dk \, k\, W(k_\pm, k) - k_0 \int \! dk\, W(k_\pm, k) \right] \nonumber \\
    &\equiv& m( k_\pm^{\rm eff} - k_0),
\end{eqnarray}
where in the penultimate line the second integral is by construction unity for properly normalized window functions, and we have defined the first integral (the centre of mass of the window function) to be $k_\pm^{\rm eff}$. Inserting this into Equation~\eqref{eq:loglike} and subsequently into Equation~\eqref{eq:k0mposterior} then provides the full posterior for $k_0$ and $m$.

Alternatively, since Equation~\eqref{eq:loglike} is a quadratic in both $m$ and $k_0$, the marginalization over $m$ can be performed analytically to give
\begin{equation}
    p(k_0| \mathbf{P}_\pm, \boldsymbol{\Sigma}, W) \propto \sqrt{\frac{2 \pi}{a}} \exp\left( \frac{b^2}{8a} \right),
\end{equation}
where
\begin{equation}
    a \equiv (\boldsymbol{\Delta k}^{\rm eff})^t \boldsymbol{\Sigma}^{-1} \boldsymbol{\Delta k}^{\rm eff}, \quad b \equiv -2 \mathbf{P}_\pm^t \boldsymbol{\Sigma}^{-1} \boldsymbol{\Delta k}^{\rm eff},
\end{equation}
with $\boldsymbol{\Delta k}^{\rm eff} \equiv ( k_-^{\rm eff} - k_0, k_+^{\rm eff} - k_0)$.

\begin{figure}[]
\includegraphics[width=0.48\textwidth]{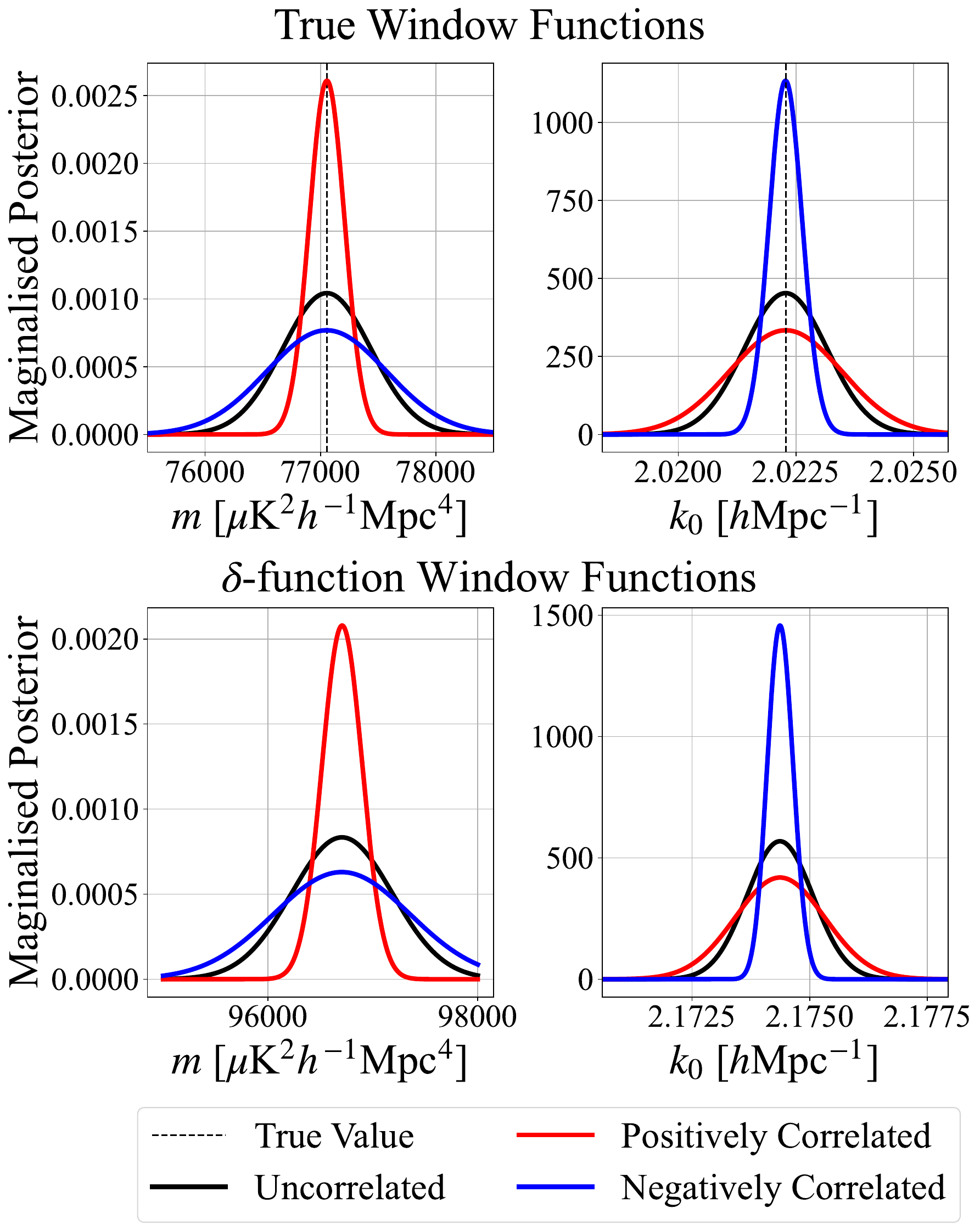}
\caption{Mock constraints on the slope $m$ and crossover scale $k_0$, expressed as marginalized posteriors where all variables except for the parameter in question have been marginalized over. Top: Constraints using the true window functions of the power spectrum estimate, as computed in Figure~\ref{fig:FW_0.01_violin}. Bottom: Biased inferences made assuming that the window functions are Dirac delta functions. Different solid lines show the effect of different error covariances while the true parameters values are shown with vertical dotted lines. One sees that neglecting window functions in analyses can lead to biased inferences, whereas the mismodeling of error covariances can lead to parameter constraints that are overly pessimistic or overly optimistic.}
\label{fig:posterior}
\end{figure}

Using Equation~\ref{eq:loglike}, we compute the marginalized posterior distributions for parameters $m$ and $k_0$ and explore various assumptions for window functions and error covariances. With the window functions we explore two different computations. In the first one, we take the correct $z=6$ window functions from Figure~\ref{fig:FW_0.01_violin} that correspond to the two data points straddling the $k_0 \approx 2\,h\textrm{Mpc}^{-1}$ crossover scale. In the second computation, we erroneously assume that the window functions are Dirac delta functions centered on the starred points in Figure~\ref{fig:FW_0.01_violin}. This is in effect what is assumed if one does not bother to compute window functions for one's power spectrum estimator. For the error covariance, we examine three possibilities. We consider a case where $P_-$ and $P_+$ are positively correlated by using the true covariance between the points (which happens to have positive off-diagonal terms in this case). Our second possibility is an uncorrelated case where the off-diagonal terms are artificially set to zero. The final possibility is the negatively correlated version where the off-diagonal terms are given the opposite sign compared to the true covariance. The final posteriors, marginalized separately for $m$ and $k_0$, are shown in Figure~\ref{fig:posterior} for all the aforementioned scenarios. Also shown (in vertical dashed lines) are the true values of the parameters inferred by acting on the true theory cross power spectra with the window functions.

Several trends are clear. Immediately, one sees the importance of including window functions in one's analyses. Without taking them into account, significant biases are present in the inference of both $m$ and $k_0$. The error covariances strongly affect the widths of the posteriors. Positive covariances give rise to narrow posteriors on the slope $m$ because coherent (positively correlated) perturbations move $P_-$ and $P_+$ up and down in concert, leaving the slope relatively unaffected. The reverse is true for inferences of $k_0$, where the two points straddling either side of the horizontal axis must be perturbed in opposite directions to keep the intercept relatively constant. Putting this together, one sees that if non-negligible covariances exist between $P_-$ and $P_+$, their mismodeling can lead to parameter inferences that can be either overconfident or underconfident. For example, in our current scenario the covariance is positive, and thus the omission of off-diagonal terms in $\boldsymbol{\Sigma}$ would result in an overconfident error bar on $k_0$.

In this section we have computed a few quantitative examples to demonstrate a qualitative point---that it is crucial to quantify one's window functions and error covariances. However, it is important not to overgeneralize our conclusions, since they can easily change depending on the subtle details of one's scenario. For example, suppose the crossover scale occurred at $k \sim 1.5\,h\textrm{Mpc}^{-1}$ at $z=7$, and one was modeling the experiment corresponding to Figure~\ref{fig:covariances}. In that case, $P_-$ and $P_+$ would be negatively correlated, leading to the opposite conclusions regarding the question of overconfidence versus underconfidence in one's inferences.

\section{Conclusion}
\label{sec:Conc}

Line-intensity mapping is rapidly becoming a promising technique for conducting efficient yet sensitive surveys of large-scale structure, with the scientific reach of potential surveys spanning both cosmological and astrophysical applications. A variety of different emission lines are being targeted by present and upcoming instruments, opening the door to a large number of possible cross-correlations. Such cross-correlations would likely represent not only a valuable scientific product, but also a means to suppress systematics such as foregrounds. However, forecasting the value of LIM cross-correlations requires the detailed characterization of hypothetical measurements and their statistical properties, quantitatively taking into account instrumental effects and the nuances of one's data analysis prescriptions.

In this paper, we present an end-to-end pipeline for simulating LIM cross-correlations. In using this pipeline to carry out a detailed forecast for upcoming \tcm-[CII] cross-correlation measurements, we explored the impact of various analysis choices and forecasting methodologies. This resulted in the following lessons learned: 
\begin{enumerate}
    \item When incoherently averaging together power spectrum estimates from $N_{\rm patches}$ different parts of the sky, the foreground residual errors average down with a $1/\sqrt{N_{\rm patches}}$ scaling (even though one might assume \emph{a priori} that foregrounds are not entirely independent in different parts of the sky).
    \item Although foreground residual errors in the power spectrum are not expected to be Gaussian distributed, some level of Gaussianization occurs in the binning step of one's power spectrum estimation. However, there typically remains some residual non-trivial behaviour in the tails of the distributions.
    \item In general, cross-terms in power spectrum variances mean that it is advisable to simultaneously vary noise, the cosmological signal, and foregrounds in one's simulations. However, in certain regimes (detailed in Section~\ref{sec:cross_terms_important}) this is reasonably similar to error estimates obtained by performing a simulation where each component is varied while others are held fixed, and then combining the resulting variations in quadrature.
    \item Error covariances between different $k$ bins of one's power spectrum estimate exist, and have a non-negligible impact when computing SNR.
    \item Window functions and error covariances should be computed and must be propagated downstream to parameter estimation. Doing so avoids parameter inference biases and the possibility of final constraints that are either underconfident or overconfident.
\end{enumerate}
We believe these lessons to be broadly applicable to a variety of forecasts using LIMs or other cosmological probes. To that end, the publicly available pipeline, \textsc{limstat}, outlined in this work can easily be adapted to not only explore cross-correlations between interferometric and single dish observations, but any combination of such measurements. For instance, one might instead be interested in understanding the cross-correlation between Ly$\alpha$ emitting galaxies detected by the Spectro-Photometer for the History of the Universe, Epoch of Reionization and Ices Explorer (SPHEREx, \citealt{SPHEREx}) and [CII] emission from CCATp. 


Cognizant of the aforementioned subtleties, we forecast the possibility of an upcoming cross-correlation measurement with HERA and CCATp, which are set to observe a small overlapping region of sky subtending 4 deg$^2$. Given this small survey area and the limited Fourier coverage of these two instruments, the commonly employed \tcm foreground mitigation strategy of excising the footprint of the so-called ``foreground wedge" is undesirable. While it is possible with such a strategy to make a marginal detection at $z = 6$ and $z = 7$, these measurements remain noise-limited. Opting to include the wedge modes in the hopes of decreasing the variance without any foreground mitigation strategy does not provide much relief. However, suppose one is able to address the \tcm foreground contaminants using other means and reduce them by a factor of 100 at the map level. Although ambitious, this requirement is still considerably easier than what is required for a \tcm auto-spectrum measurement. If such a level of mitigation can be achieved, we find that the cross-correlations can be a huge benefit in teasing out a 6.79$\sigma$ detection at $z = 6$ and a 6.84$\sigma$ detection at $z = 7$. Going to higher redshifts remains a challenge since both the noise and the foreground spectra brighten significantly with decreasing frequency. Nonetheless, these forecasts indicate that this first generation of \tcm-[CII] cross-correlations may help us rule our certain reionization scenarios which produce ionization bubbles on very large scales. That being said, much of the parameter space of possible cross-over scales is not probed by the poor Fourier coverage of these instruments. 

Given these limitations, we look to the future and provide a forecast for a set of hypothetical new instruments. We go to larger survey areas and also achieve better Fourier coverage by improving the spectral resolution of CCATp. This allows one to make measurements of higher statistical significance, as well as to cover a greater variety of possible theoretical models (which for example often make different predictions for the crossover scale at which \tcm $\times$ [CII] correlations transition from negative to positive). Over increased survey areas of 40 deg$^2$ and $400\,\textrm{deg}^2$, we once again find that the most successful strategy is to include wedge modes and mitigate the \tcm foregrounds to roughly the percent level. Over the larger survey area of $400\,\textrm{deg}^2$ precision measurements can be made across the cross-over scale, thus potentially providing constraints on both the timing and morphology of reionization. 

While this pipeline has allowed us to perform a relatively realistic forecast, there remain a myriad of additional systematics that were not considered. Currently, \tcm observations are challenged by a variety of systematics including, but not limited to, radio frequency interference \citep{Wilensky_2019_RFI}, ionospheric effects \citep{Jeli_2010_ionosphere,Moore_2013_ionosphere,Martinot_2018_ionosphere}, calibration errors \citep{Orosz_2019_calibration,Byrne_2019_calibration,Barry_2016_calibration}, and mutual coupling \citep{Rath_Pascua_2024}. While only a small number of sub-mm CO and [CII] mapping instruments have come online, we have already been able to learn a great deal about the systematic contaminants that may be detrimental to these types of surveys. Side lobe pickup from the ground \citep{Foss_2022_COMAP}, atmospheric variability associated with Lissajous scanning strategies \citep{Ihle_2022_COMAP}, and mechanical degradation \citep{Lunde_2024_COMAP} have all shown to decrease data quality. It is still unclear how these systematics will affect a cross-correlation measurement. 

To conclude, this work underscores the significant potential of LIM-LIM cross-correlations in constraining the EoR in the near future. Using our simulation and analysis pipeline, we have elucidated the importance of end-to-end modeling and the need for the rigorous quantification of statistical properties. Our forecasts for upcoming HERA and CCATp cross-correlation measurements demonstrate both the challenges and opportunities ahead. Moving forward, it is crucial to pursue unified efforts among various experiments and coordinate common survey areas. By fostering collaborative initiatives, we maximize scientific returns and inch closer to achieving our shared vision of understanding the enigmatic early stages of structure formation.

\begin{acknowledgments}
HF would like to thank Adélie Gorce, Robert Pascua, Rebecca Ceppas de Castro, and Kai-Feng Chen for being the first users of this pipeline and acknowledges their work in co-authoring this piece of software, greatly expanding its functionality and user-friendliness. The authors would also like to thank Simon Foreman, Jessica Avva Zebrowski, Kirit Karkare, Cathryn Trott, Anirban Roy, Kana Moriwaki, Miguel Morales, Bryna Hazelton, Steven Murray, Anthony Pullen, Patrick Breysse, Jordan Mirocha, Saurabh Singh, and Michel Adamič for many insightful discussions. 

HF is supported by the Fonds de recherche du Québec Nature et Technologies (FRQNT) Doctoral Research Scholarship award number 315907. AL and HF acknowledge support from the Trottier Space Institute, the New Frontiers in Research Fund Exploration grant program, an FRQNT New University Researchers Grant, the Canadian Institute for Advanced Research (CIFAR) Azrieli Global Scholars program, a Natural Sciences and Engineering Research Council of Canada (NSERC) Discovery Grant and a Discovery Launch Supplement, the Sloan Research Fellowship, and the William Dawson Scholarship at McGill. This research was enabled in part by support provided by Compute Canada (\href{www.computecanada.ca}{www.computecanada.ca}).

\end{acknowledgments}


\vspace{5mm}

\newpage
\appendix

\section{The Statistical Modeling Of Correlated Fields}\label{append:correlation}

In \citet{Pagano_Liu2020}, a simple statistical model for decorrelating fields was presented in the context of varying the degree of correlation between the ionization and density fields during reionization. This model generated fields that produce the same degree of correlation on all length scales. Here we extend this model to generate fields with scale-dependent correlations.

Suppose we are interested in generating two fields $T_a(\mathbf{r})$ and $T_b(\mathbf{r})$ such that they are consistent with having power spectra $P_a(k)$ and $P_b(k)$, respectively, i.e.,
\begin{equation}
    \langle \tilde{T}_a (\mathbf{k}) \tilde{T}_a (\mathbf{k}^\prime)^* \rangle = (2\pi)^3 \delta^D (\mathbf{k} - \mathbf{k}^\prime) P_a (k) \quad \textrm{and} \quad  \langle \tilde{T}_b (\mathbf{k}) \tilde{T}_b (\mathbf{k}^\prime)^* \rangle = (2\pi)^3 \delta^D (\mathbf{k} - \mathbf{k}^\prime) P_b (k),
    \label{eq:doublepspecdef}
\end{equation}
where $\delta^D$ denotes a Dirac delta function and we adopt the standard cosmological Fourier convention, such that
\begin{equation}
    \tilde{T}(\mathbf{k}) \equiv \int \! d^3 r \, e^{-i \mathbf{k}\cdot \mathbf{r}} T(\mathbf{r}) \quad \textrm{and} \quad  T(\mathbf{r}) = \int \! \frac{d^3 k}{(2\pi)^3} \, e^{i \mathbf{k}\cdot \mathbf{r}} \tilde{T}(\mathbf{k}).
\end{equation}

To ensure that the our two fields obey not only the right autocorrelation statistics but also have the right cross power spectrum, we generate the fields one at a time. For $T_a(\mathbf{r})$ we simply follow the usual procedure and draw random realizations in Fourier space with variance consistent with Equation~\eqref{eq:doublepspecdef}. The second field is then created by computing
\begin{equation}\label{eq:correlated_fields}
\tilde{T}_b(\mathbf{k})  = \tilde{T}_a(\mathbf{k}) f(k) e^{-i\phi}.
\end{equation}
The factor $f(k)$ adjusts the power of field $b$ such that it has the auto-spectrum $P_b(k)$. Computing the two-point function of both sides of Equation~\eqref{eq:correlated_fields} reveals that
\begin{equation}
    P_b(k) = P_a(k) f^2 (k) \quad \Rightarrow \quad f(k) = \sqrt{\frac{P_b(k)}{P_a(k)}},
\end{equation}
giving us the $f(k)$ required to satisfy our constraints on the auto spectra.

The factor $e^{-i\phi}$ adjusts the relative phases of the fields and ensures that the cross spectrum of fields $a$ and $b$ can be tuned to a desired $P_{ab}(k)$. In \cite{Pagano_Liu2020}, the values of $\phi$ were drawn from a Gaussian distribution with mean zero and standard deviation $\sigma$ for all comoving wavenumber, $k$.

Here we seek to solve for $\sigma$ as a function of $k$ by relating it to known $k$-dependent quantities, the auto- and cross-spectrum. Forming the cross power spectrum between our two fields gives
\begin{equation}
\label{eq:crosscorralgebra}
    P_{ab}(k) \propto \langle \tilde{T}_a(k) \tilde{T}^*_b(k) \rangle = f(k) \langle e^{-i\phi} \rangle \langle \tilde{T}_a(k) \tilde{T}^*_a(k) \rangle \propto f(k) \langle e^{-i\phi} \rangle P_a(k) = \langle e^{-i\phi} \rangle \sqrt{P_a(k) P_b(k)}
\end{equation}
If $\phi$ is drawn from a Gaussian distribution, its expectation value can be readily evaluated to give
\begin{equation}
\label{eq:eiphiexpectation}
\langle e^{-i\phi} \rangle = \int P(\phi) e^{-i\phi} = \frac{1}{ \sqrt{2\pi \sigma^2}}\int\! d\phi \, e^{-\phi^2 / 2\sigma^2} e^{-i\phi} = \exp \left[ \frac{\sigma^2 (k)}{2} \right],
\end{equation}
where in the last expression we have explicitly included the potential $k$ dependence of $\sigma$. It is precisely this $k$ dependence that enables one to engineer a scale-dependence on the degree of correlation between our two fields. Comparing Equations~\eqref{eq:crosscorralgebra} and \eqref{eq:eiphiexpectation}, one sees that getting the right cross-correlation power spectrum $P_{ab}(k)$ can be achieved by setting
\begin{equation}
    \label{eq:complexsigma}
    \frac{\sigma^2}{2} \equiv \ln |r(k)| + \frac{\pi}{2} \left( 1 - \rm{sgn}[r(k)]\right),
\end{equation}
where $\rm{sgn}$ is the sign function. This is equivalent to Equation~\eqref{eq:r(k)_sigma} given in Section~\ref{sec:cosmo_fields}, where one simply sets $\sigma(k) = \sqrt{\ln(|r(k)|^{-2})}$ and multiplies the right hand side of Equation~\eqref{eq:correlated_fields} by $\textrm{sgn}[r(k)]$.

In summary, drawing random phases for each location in $\textbf{k}$ according to Equation~\eqref{eq:complexsigma} and then evaluating Equation~\eqref{eq:complexsigma} gives two fields with the desired auto- and cross-correlation spectra.

\section{Power Spectrum Window Functions}
\label{sec:windowfunctionderiv}
In this appendix we derive expressions that relate the true cross-correlation power spectrum $P_{ab}(k)$ to the measured power spectrum $\hat{P}_{ab}(k)$ that is distorted by one's instrument response. Our goal is to find a window function $W_{ab}(k, k^\prime)$ such that
\begin{equation}
\label{eq:1Dwindowdef}
\langle \hat{P}_{ab}(k) \rangle = \int \! dk^\prime W_{ab}(k, k^\prime) P_{ab}(k^\prime).
\end{equation}
To do so, we first write the observed sky map of an instrument $T^{\rm obs} (\mathbf{r})$ as
\begin{equation}
\label{eq:Tobs}
T^{\rm obs} (\mathbf{r}) = \int \! d^2 u \exp \left( i 2 \pi \frac{\mathbf{u} \cdot \mathbf{r}_\perp}{r_z} \right) \tilde{G} (\mathbf{u}, \nu_z) \int d^2 \! \phi \, \exp(- i 2 \pi \mathbf{u} \cdot \boldsymbol{\phi}) A (\boldsymbol{\phi}, \nu_z) T(\boldsymbol{\phi}, \nu_z),
\end{equation}
where we are working in the flat sky limit given the (relatively) narrow fields considered in this paper. This allows us to identify a near-unambiguous line of sight direction that we can align with the $z$ direction of our coordinate system, leaving the $x$ and $y$ directions to span the transverse space perpendicular to the line of sight, such that $\mathbf{r}_\perp \equiv (r_x, r_y)$ . Again utilizing the flat-sky limit, we can write $\mathbf{r} \equiv (r_x, r_y, r_z) = (\mathbf{r}_\perp, r_z) = (r_z \boldsymbol{\phi}, r_z)$. Here, $r_z$ is the radial comoving distance and $\boldsymbol{\phi} \equiv (\phi_x, \phi_y)$ refers to small angles corresponding to $r_x$ and $r_y$. We denote the frequency of observation as $\nu_z$, with the subscript ``$z$" to remind ourselves of the one-to-one correspondence between frequency and radial distance in intensity mapping. In words, Equation~\eqref{eq:Tobs} says that the sky is multiplied by some instrumental configuration space response $A(\boldsymbol{\phi}, \nu_z)$, then Fourier transformed in the transverse direction before being sampled by an instrumental Fourier space response $\tilde{G}$ (which will often be frequency dependent) and Fourier transformed back to configuration space.

Although we deal with two reasonably different types of instruments in this paper (interferometers and single dish telescopes), Equation~\eqref{eq:Tobs} is general enough to accommodate both. For example, for a radio interferometer $A(\boldsymbol{\phi}, \nu_z)$ would represent the primary beam and $\tilde{G}(\mathbf{u}, \nu_z)$ would encode the sampling of the $uv$-plane by the baselines of the interferometer (hence the suggestive use of $\mathbf{u}$ as the Fourier dual to sky angle $\boldsymbol{\phi}$). For a single dish telescope, one might set $A(\boldsymbol{\phi}, \nu_z)$ to unity, whereas $\tilde{G}(\mathbf{u}, \nu_z)$ would be the Fourier transform of the telescope's point spread function.

To estimate the power spectrum, the data analyst can (optionally) multiply by some tapering/apodization function $B(\mathbf{r})$, defining $T^{\rm tap} (\mathbf{r}) \equiv B (\mathbf{r}) T^{\rm obs}(\mathbf{r})$. This is then Fourier transformed to obtain
\begin{equation}
\tilde{T}^{\rm tap} (\mathbf{k}) \equiv \int \! d^3 r \, e^{-i \mathbf{k} \cdot \mathbf{r}} T^{\rm tap} (\mathbf{r}) =  \int \frac{d^3 k^\prime}{(2\pi)^3} \tilde{T}(\mathbf{k}^\prime) F(\mathbf{k}, \mathbf{k}^\prime),
\end{equation}
where
\begin{eqnarray}
\label{eq:maplevelwindow}
F(\mathbf{k}, \mathbf{k}^\prime) &\equiv & \int \! d^3 r \, e^{-i \mathbf{k}\cdot \mathbf{r}} e^{i k_z^\prime r_z} B(\mathbf{r}) \int \! d^2 u \, e^{i 2 \pi \mathbf{u} \cdot \mathbf{r}_\perp / r_z} \tilde{G} (\mathbf{u}, \nu_z) \int \! d^2 \theta A(\boldsymbol{\theta}, \nu_z) e^{- i (2\pi \mathbf{u} - \mathbf{k}_\perp^\prime r_z) \cdot \boldsymbol{\theta}} \nonumber \\
&=& \int \! dr_z \! \int \! d^2 \mathbf{u} \,e^{-i (k_z - k_z^\prime) r_z} \tilde{B} \left( \mathbf{k}_\perp - \frac{2 \pi \mathbf{u}}{r_z}, r_z \right) \tilde{G} (\mathbf{u}, \nu_z) \tilde{A} \left( \mathbf{u} - \frac{\mathbf{k_\perp^\prime} r_z}{2 \pi}, \nu_z \right),
\end{eqnarray}
with $\mathbf{k}_\perp \equiv (k_x, k_y)$ as the Fourier dual to $\mathbf{r}_\perp$. We note that we are mixing two different Fourier conventions here: for quantities that mostly pertain to an instrumental response, we use the Fourier convention where a factor of $2\pi$ appears in the exponent and the forward and backward transforms differ only by a sign in the exponent; for cosmological fields and $B(\mathbf{r})$ we adopt the standard cosmological convention stated in Appendix~\ref{append:correlation}.

Forming an estimator $\hat{P}_{ab}(\mathbf{k})$ for the cross power spectrum of two surveys involves cross-correlating two copies of $\tilde{T}_{\rm obs} (\mathbf{k})$, one from each instrument, to give
\begin{equation}
\label{eq:phat3Ddef}
\hat{P}_{ab}(\mathbf{k}) \equiv \frac{\langle \tilde{T}^{\rm tap}_{a} (\mathbf{k}) \tilde{T}^{\rm tap}_{b} (\mathbf{k})^* \rangle }{V N(\mathbf{k})} = \frac{1}{V N(\mathbf{k})} \int \!\frac{d^3 k^\prime}{(2\pi)^3} \, P_{ab} (k^\prime) F_a (\mathbf{k}, \mathbf{k}^\prime) F_b (\mathbf{k}, \mathbf{k}^\prime)^* = \int \! d^3 k^\prime \,  P_{ab} (k^\prime) W_{ab}^{\rm 3D} (\mathbf{k} ,\mathbf{k}^\prime),
\end{equation}
where $N(\mathbf{k})$ is a normalization factor (to be derived later), and $V$ is the overlap volume of our two surveys, which are labelled using subscripts $a$ and $b$ for each of the two surveys. In the second equality we used the definition of the power spectrum, namely $\langle \tilde{T}_a (\mathbf{k}) \tilde{T}_b (\mathbf{k})^* \rangle = (2\pi)^3 \delta^D (\mathbf{k}- \mathbf{k}^\prime) P(k)$, where $\delta^D$ signifies a Dirac delta function. In the final equality, we defined a window function $W_{ab}^{\rm 3D}$ that is analogous to the one defined in Equation~\eqref{eq:1Dwindowdef}, except in three-dimensional Fourier space. This version of the window function captures the way in which the estimated power spectrum $\hat{P}_{ab}(\mathbf{k})$ can be anisotropic in $\mathbf{k}$ thanks to our survey instruments, even though we have assumed that the true power spectrum $P_{ab} (k)$ is statistical isotropic and depends only on $k$.

Deriving explicit expressions for $W_{ab}^{\rm 3D}$ is important for two reasons. First, it allows us to verify that our power spectra are correctly normalized by enforcing that $\int \! d^3 k^\prime \, W_{ab}^{\rm 3D} (\mathbf{k}, \mathbf{k}^\prime) = 1$. In practice, this can be done by reverse engineering, where one imposes the normalization condition in order to derive the normalization $N(\mathbf{k})$, which is given by
\begin{equation}
N(\mathbf{k}) \equiv  \int \! d^3 k^\prime \,   W_{ab}^{\rm 3D}(\mathbf{k} ,\mathbf{k}^\prime) = \frac{1}{V} \int \!\frac{d^3 k^\prime}{(2\pi)^3} \,  F_a (\mathbf{k}, \mathbf{k}^\prime) F_b (\mathbf{k}, \mathbf{k}^\prime)^*.
\end{equation}
The second reason for computing $W_{ab}^{\rm 3D} (\mathbf{k} ,\mathbf{k}^\prime)$ is to eventually bin it to yield $W_{ab} (k, k^\prime)$, which determines the width of horizontal error bars in our final power spectrum estimates.

Consider first the issue of normalization. To simplify the expressions, we will assume that $B(\mathbf{r})$ is a function of frequency only. This is not an approximation, but simply a reflection of how we use $B(\mathbf{r})$ in this paper: as a spectral apodization function to avoid edge effects from bright foregrounds abruptly dropping to zero outside the survey volume. With this assumption, the spatial part of $\tilde{B}$ is proportional to a Dirac delta function, and one obtains
\begin{equation}
F(\mathbf{k}, \mathbf{k}^\prime) \approx  \int \! dr_z \, e^{-i (k_z - k_z^\prime) r_z} r_z^2 B(r_z)   \tilde{G} \left(\frac{\mathbf{k}_\perp r_z}{2\pi}, \nu_z \right) \tilde{A} \left( \frac{\mathbf{k_\perp} r_z}{2 \pi} - \frac{\mathbf{k_\perp^\prime} r_z}{2 \pi}, \nu_z \right),
\end{equation}
which eventually yields
\begin{equation}
N(\mathbf{k}) = \frac{1}{V} \int \! dr_z\,  \tilde{G}_a \left(\frac{\mathbf{k}_\perp r_z}{2\pi}, \nu_z \right) \tilde{G}_b \left(\frac{\mathbf{k}_\perp r_z}{2\pi}, \nu_z \right)^* B_a (r_z) B_b (r_z) r_z^2 \int \! d^2 \theta \,A_a (\boldsymbol{\theta}, \nu_z) A_b (\boldsymbol{\theta}, \nu_z).
\end{equation}
For transverse modes on larger lengthscales than the resolution of one's instrument, we have $\tilde{G} \rightarrow 1$, and this normalization will be close to order unity because the integrals reduces approximately to the survey volume (cancelling out the $1/V$ term), with just a slight modification for the spatial response $A(\boldsymbol{\theta}, \nu_z)$ and the frequency taper $B(r_z)$.

With a correctly normalized power spectrum estimator, the only interesting information in a window function is its shape. This is given by
\begin{eqnarray}
\label{eq:raw3Dwindow}
W_{ab}^{\rm 3D} (\mathbf{k} ,\mathbf{k}^\prime) &=& \int \frac{dr_z dr_z^\prime}{V (2\pi)^3} e^{-i(k_z-k_z^\prime) r_z} e^{i (k_z - k_z^\prime) r_z^\prime} r_z^2 r_z^{\prime 2} B_a (r_z) B_b (r_z^\prime) \tilde{G}_a \left(\frac{\mathbf{k}_\perp r_z}{2\pi}, \nu_z \right) \tilde{G}_b \left(\frac{\mathbf{k}_\perp r_z^\prime}{2\pi}, \nu_z^\prime \right)^* \nonumber \\
&& \times \tilde{A}_a \left( \frac{\mathbf{k_\perp} r_z}{2 \pi} - \frac{\mathbf{k_\perp^\prime} r_z}{2 \pi}, \nu_z \right) \tilde{A}_b \left( \frac{\mathbf{k_\perp} r_z^\prime}{2 \pi} - \frac{\mathbf{k_\perp^\prime} r_z^\prime}{2 \pi}, \nu_z^\prime \right)^*.
\end{eqnarray}
While it is exact (up to the flat-sky approximation), this expression is unfortunately computational infeasible to use in practice. Instead, we must compute the binned version of the window function analytically. This can be done by making a few simplifying assumptions. First, we will take $A(\boldsymbol{\phi}, \nu_z)$ to be a frequency-independent quantity for our interferometer. In general, this will not be the case for a real observation. However, recall from Section~\ref{sec:21cmFGs} that our analysis involves carving out just a small central portion of our interferometer's large field-of-view, making a frequency-independent treatment more justifiable. Second, whenever $r_z$ appears because it is used as a conversion factor between angles and transverse distances, we replace it with $D_c$, which we define as $r_z$ evaluated at the midpoint of our survey volume. With both of these approximations we are essentially treating slowly varying functions of frequency as being perfectly constant. In contrast, we certainly cannot do the same for $T(\boldsymbol{\phi}, \nu_z)$ or $\tilde{G}(\mathbf{u}, \nu_z)$. (The latter's frequency dependence, for instance, is responsible for the phenomenology of the foreground wedge in interferometeric measurements; \citealt{2014PhRvD..90b3018L}). Finally, we will assume that the response of our instruments are azimuthally symmetric (precluding the possibility of elliptical point spread functions, for example). This enables us to take $\tilde{G}(\mathbf{u}, \nu_z)$ to $\tilde{G}(|\mathbf{u}|, \nu_z)$.

To implement our approximation of a narrow field cut out from a wide field, we imagine that both $A_a(\boldsymbol{\theta})$ and $A_b (\boldsymbol{\theta})$ take Gaussian forms with standard deviations $\theta_a$ and $\theta_b$, respectively. The last two terms of Equation~\eqref{eq:raw3Dwindow} then become
\begin{equation}
    \tilde{A}_a (\mathbf{q}) \tilde{A}_b (\mathbf{q})^* \propto 2 \pi^2 (\theta_a^2 + \theta_b^2) \exp \left[ 2 \pi^2 (\theta_a^2 + \theta_b^2) |\mathbf{q}|^2 \right] \rightarrow \delta^D (\mathbf{q}),
\end{equation}
where $\mathbf{q} \equiv (\mathbf{k}_\perp - \mathbf{k}_\perp^\prime) D_c / 2 \pi$ and in the last step we took the limit $\theta_a, \theta_b \rightarrow \infty$ to encode the flat response in the field of interest. (Essentially, we are taking the very flat top portion of a Gaussian). Inserting this into Equation~\eqref{eq:raw3Dwindow} and imposing our azimuthally symmetric approximation on $\tilde{G}$ then gives
\begin{equation}
W_{ab}^{\rm 3D} (\mathbf{k} ,\mathbf{k}^\prime) \propto \delta^D (\mathbf{k}_\perp - \mathbf{k}_\perp^\prime) W^\parallel_{ab} (k_\perp, k_z - k_z^\prime),
\end{equation}
where
\begin{equation}
W^\parallel_{ab} (k_\perp, k_z ) \equiv F^\parallel_a (k_\perp, k_z ) F^\parallel_b (k_\perp, k_z)^*,
\end{equation}
with $k_\perp \equiv | \mathbf{k}_\perp |$, and
\begin{equation}
F^\parallel (k_\perp, k_z) \equiv \int dr_z e^{-i k_z r_z} B(r_z) \tilde{G} \left(\frac{k_\perp D_c}{2\pi}, \nu_z \right),
\end{equation}
with $a$ and $b$ subscripts adorning each term as appropriate. At this point, we have arrived at a computationally tractable form $W_{ab}^{\rm 3D} (\mathbf{k} ,\mathbf{k}^\prime)$, our approximations having resulted in symmetries that reduce the complexity of the expressions (such as the fact that only the \emph{difference} between $k_z$ and $k_z^\prime$ matters).

To move towards a final expression for the binned window function $W_{ab} (k, k^\prime)$, we first bin azimuthally in the $k_x k_y$ plane down to $k_\perp$. Defining $\varphi_k$ as the azimuthal angle within a cylindrical coordinate system defined by $k_\perp$, $\varphi_k$, and $k_z$ (and similarly for primed coordinates), we can define
\begin{equation}
W^{\rm 2D}_{ab} (k_\perp, k_z; k_\perp^\prime, k_z^\prime) \propto \int d \varphi_k d\varphi_k^\prime W^{\rm 3D}_{ab} ( \mathbf{k}, \mathbf{k}^\prime),
\end{equation}
and using the fact that $\delta^D (\mathbf{k}_\perp - \mathbf{k}_\perp^\prime) = \delta^D (k_\perp - k_\perp^\prime) \delta^D (\varphi_k - \varphi_k^\prime) / k_\perp$, we have
\begin{equation}
W^{\rm 2D}_{ab} (k_\perp, k_z; k_\perp^\prime, k_z^\prime) \propto \delta^D (k_\perp - k_\perp^\prime) W^\parallel_{ab} (k_\perp, k_z - k_z^\prime).
\end{equation}
From here, there are two further binning operations that one can perform. The first is relatively straightforward, which is to fold $\pm k_z$ and $\pm k_z^\prime$ into always-positive $k_\parallel$ and $k_\parallel^\prime$ values. The second is to bin in circular $k = \sqrt{k_\perp^2 + k_\parallel^2}$ rings. This step is best done numerically---unlike the azimuthal binning that considerably simplified our expressions by taking advantage of azimuthal symmetries (or approximate symmetries) in our survey instruments, spherical binning has no such symmetry to leverage, since intensity mapping surveys probe line-of-sight fluctuations in a different way than they do transverse fluctuations.

\bibliography{bibliography}{}
\bibliographystyle{aasjournal}

\end{document}